\documentclass{aastex62} 

\received{June 3, 2018}
\revised{October, 2018}
\accepted{November, 2018}
\submitjournal{PASP Special Volume on Machine Learning}

\shorttitle{Deep Learning Applied to the Asteroseismic Modeling of Stars}
\shortauthors{Hendriks \& Aerts}

\begin{document} 

\title{Deep Learning Applied to the Asteroseismic Modeling of Stars with
  Coherent Oscillation Modes}

\correspondingauthor{Luc Hendriks}
\email{luc.hendriks@gmail.com}

\author{L.\ Hendriks}
\affiliation{Institute for Mathematics, Astrophysics and Particle Physics,
  Faculty of Science, Mailbox 79, Radboud University Nijmegen, P.O. Box 9010,
  NL-6500 GL Nijmegen, The Netherlands}

\author{C.\ Aerts}
\affiliation{Institute of Astronomy, KU\,Leuven, Celestijnenlaan 200D,
B-3001 Leuven, Belgium}
\affiliation{Institute for Mathematics, Astrophysics and Particle Physics,
  Faculty of Science, Mailbox 79, Radboud University Nijmegen, P.O. Box 9010,
  NL-6500 GL Nijmegen, The Netherlands}

\begin{abstract}
  We develop a novel method based on machine learning principles to achieve
  optimal initiation of CPU-intensive computations for forward asteroseismic
  modeling in a multi-D parameter space.  A deep neural network is trained on a
  precomputed asteroseismology grid containing about 62 million coherent
  oscillation-mode frequencies derived from stellar evolution models. These
  models are representative of the core-hydrogen burning stage of
  intermediate-mass and high-mass stars. The evolution models constitute a 6D
  parameter space and their predicted low-degree pressure- and gravity-mode
  oscillations are scanned, using a genetic algorithm. A software pipeline is
  created to find the best fitting stellar parameters for a given set of
  observed oscillation frequencies. The proposed method finds the optimal
  regions in the 6D parameters space in less than a minute, hence providing the optimal
  starting point for further and more detailed forward asteroseismic modeling in
  a high-dimensional context. We test and apply the method to seven pulsating
  stars that were previously modeled asteroseismically by classical grid-based
  forward modeling based on a $\chi^2$ statistic and obtain good agreement with
  past results. Our deep learning methodology opens up the application of
  asteroseismic modeling in +6D parameter space for thousands of stars pulsating
  in coherent modes with long lifetimes observed by the {\it Kepler\/} space
  telescope and to be discovered with the TESS and PLATO space missions, while
  applications so far were done star-by-star for only a handful of cases. Our method is open source and can be used by anyone freely\footnote{The source code to run the analysis can be found at \tt{https://github.com/l-hendriks/asteroseismology-dnn}.}
\end{abstract}


\section{Introduction}

The past decade has seen tremendous progress in stellar physics, largely thanks
to the application of asteroseismology \citep{Aerts2010}. Such applications are
nowadays based on long-term uninterrupted space photometry with a precision of
$\mu\,$mag assembled for thousands of stars by the European CoRoT
\citep{Auvergne2009} and the NASA {\it Kepler\/} \citep{Koch2010} space
telescopes. This data offers the opportunity to confront stellar evolution
theory with asteroseismic data probing the stellar interior, rather than with
spectroscopic data giving information on the stellar surface.

Comparison between asteroseismic data and stellar models is done by computing
the spectrum of oscillation modes predicted by the theoretical models.
Oscillation modes are called pressure modes or gravity modes, depending on
whether the pressure force, respectively buoyancy, is the dominant restoring
force.  Pressure modes probe the physical conditions in the stellar envelope
while gravity modes tune the deep stellar interior.  A modern educational tutorial
containing various statistical methods for asteroseismology is available in
\citet{Appourchaux2014}.  

Statistical inference in asteroseismology can conveniently be done from forward
modeling. Usually, one adopts a grid-based approach and considers millions of
stellar structure models of different evolutionary stage.  Applications to
low-mass stars in the core hydrogen burning stage can be found in the pioneering
papers by \citet[][]{Miglio2005} and \citet{Quirion2010}, while
\citet{Brassard2001} developed methodology to treat coherent gravity modes in
core helium burning subdwarf stars.  In the past few years, forward
modeling based on space asteroseismology has been focused strongly on solar-like
oscillations in low-mass stars because the {\it Kepler\/} mission provided data
for thousands such objects. Their damped pressure modes are stochastically
excited by the turbulent convection in their outer convective envelope and have
periodicities ranging from minutes for the core hydrogen burning phase to hours
for red giants. The forward modeling for such ``sun-like'' stars relies on
stellar models whose input physics is taken to be very similar to the one of
solar models calibrated from helioseismology.  Such applications are nowadays often
based on Bayesian statistics, e.g.\ as in
\citet{Appourchaux2012b,Gruberbauer2013,Corsaro2014,Davies2016,Deheuvels2016,SilvaAguirre2017},
among others.

A prominent application of Machine Learning (ML) applied to low-mass star
asteroseismology based on damped pressure modes was developed by
\citet{Bellinger2016}. It is based on the combined use of spectroscopic and
asteroseismic input and relies on the principle of random forests.  This
application covers the mass range $[0.7,1.6]$\,M$_\odot$ and delivers parameter
estimation in a 7D space: stellar mass, age, initial helium, initial metalicity,
mixing length value, convective overshoot, and an amplification factor for
atomic diffusion (excluding radiative levitation, i.e., focusing only on the
settling of elements heavier than hydrogen). The metalicity and effective
temperature were found to be the two dominant features feeding the ML
code. These turn out to be about a factor three higher in feature importance
than the seismic large and small frequency separations and frequency ratios.

The situation is totally different in our application to stars of higher
  mass. The required spectroscopic input is not available at sufficiently high
  precision for those. Moreover, one cannot use scaled versions of solar physics
  because ingredients that do not occur in low-mass stars dominate for
  intermediate- and high-mass 
  stars.  In particular, the initial helium content and settling of heavy
  elements play inferior roles compared to core overshooting and macroscopic
  chemical mixing due to various phenomena like rotation and waves in the
  radiative envelope.  From a seismic point of view, intermediate- and high-mass
  stars born with a convective core and a radiative envelope
  ($M\gtrsim 2\,$M$_\odot$) reveal coherent modes with long lifetimes, rather
  than damped modes in low-mass stars. The frequencies of coherent 
modes can be
  fitted directly, because they are not subject to uncertain physics in the
  stellar atmosphere as occurs for damped modes triggered by envelope
  convection.  Due to these differences in the treatment of the theory of
  stellar structure and of stellar oscillations for the different mass regimes,
  their asteroseismic modeling requires different approaches.  \citet[][hereafter
  termed Paper\,I]{Paper-I} provided an extensive discussion on the physical and
  pulsational ingredients that are dominant for asteroseismic modeling of stars
  born with a convective core. Applications in the recent literature
\citep{Kurtz2014,Saio2015,Moravveji2015,Moravveji2016,Murphy2016,SchmidAerts2016,VanReeth2016,Sowicka2017,Kallinger2017,Szewczuk2018}
were done at a very different level of detail 
compared to low-mass pulsators and, with the exception of
  \citet{Johnston2018},  none so far achieved error estimation taking
  into account parameter correlations and the fact that the stellar models
  depend nonlinearly on the parameters. 

The mass of core hydrogen burning stars with detected coherent modes ranges
roughly from 2 to 30 solar masses.  For such stars, the physics of the near-core
region is largely unknown.  Moreover, mass loss due to a radiation-driven wind
is active as of stellar birth for masses above $\sim 20\,$M$_\odot$ and for
those a high binary rate occurs.  One therefore faces tremendous challenges in
the forward modeling of these stars in terms of dimensionality and CPU
requirements compared to the ``easy'' case of low-mass sun-like pulsators.
On the other hand, their coherent modes, and
in particular gravity modes, probe the deepest layers of those stars with high
precision, rather than just the envelope as damped pressure modes
in low-mass stars. It has been shown that  this 
property offers the opportunity to test the near-core physics
  \citep{Pedersen2018}. Gravity-mode asteroseismology opens up the possibility
to derive the helium core mass near the end of the core hydrogen burning stage
from such modes, a prominent quantity that lays the foundation for the further
evolution of the star in terms of nuclear yields. 

The current theory of angular momentum transport fails to explain
  asteroseismic data by more than an order of magnitude \citep[see][for an
  extensive review of the current status of interior rotation rates of more than
  a thousand stars]{Aerts2019}.  This theory can be improved from gravity-mode
  asteroseismology by applying the method in Paper\,I.  Similarly, the theory of
  element transport can now be calibrated asteroseismically \citep[see][for a
  review]{Salaris2017}. The potential gain from asteroseismology of high-mass
stars is thus much larger than for low-mass stars in terms of chemical
enrichment of the galaxy.

Here, we focus on intermediate- and high-mass core hydrogen burning stars in the
mass range $[2,20]\,$M$_\odot$ with detected coherent oscillation modes.  Our
aim is to develop and provide an optimal deep learning tool as a critical
start-up ingredient for efficient forward asteroseismic modeling in a +6D
parameter space. In particular, we wish to explore the performance of a
  deep learning network in order to gain computational efforts, because the
  ``manual'' asteroseismic modeling applied to a few stars with coherent modes
  so far required months of analysis work, with a high risk of missing local
  minima in parameter space \citep[cf.\ ][for an example of such a systematic
  efforts]{Ausseloos2004}.  This is not practical for future applications to
  samples of tens to hundreds of stars anticipated by future space missions,
  such as TESS \citep{Ricker2016} and PLATO \citep{Rauer2014}.  In this paper,
we restrict to slowly-rotating pulsators. Future work will consider deep
learning methods for asteroseismology of fast rotators.

\section{Stellar Models and Their Coherent Oscillations}

Aside from other variability features, a large fraction of the core-hydrogen
burning intermediate-mass and high-mass stars reveal so-called coherent
modes. In contrast to the damped oscillation modes of low-mass stars, these
coherent modes are self-driven by a heat mechanism and have long lifetimes of
thousands to millions of years \citep[e.g.,][Chapter\,2]{Aerts2010}. This
implies that these modes are easily detectable in high-precision uninterrupted
long-duration time-series data, provided that they are excited with amplitudes
above the detection threshold of the instrument.  The time base of such data
is crucial because, unlike the short periods of the damped oscillations of
low-mass stars, the coherent modes we are dealing
with here have periods from about an hour up to several days, leading to
years-long beating patterns to be covered with data. This is particularly the case for
gravity-mode oscillations, which probe the deep stellar interior.  As such, the
4-year nominal {\it Kepler\/} mission provides the best observations to analyze
the period-spacing patterns expected for such modes from theory
\citep{Miglio2008,Bouabid2013} and actually detected in data 
\citep[e.g.,][and references
therein]{VanReeth2015,Bowman2017,Papics2017}. Indeed, while the discovery of
period-spacing patterns of gravity modes could be achieved from the 150-day
light curves assembled with CoRoT \citep{Degroote2010,Papics2012}, this time
base was too short to unravel all the beating patterns between the modes and to
identify them in terms of their degree $l$ and azimuthal order $m$.

The 4-year long monitoring from space by {\it Kepler\/} revealed suitable data
sets to detect and identify the degree and radial order, $(l,n_{pg})$, 
of several to tens of coherent gravity modes,
opening up the opportunity of forward asteroseismic modeling of such modes.  As
said, we focus on stars born with a convective core and a radiative envelope, in
the core hydrogen burning phase of their evolution (the so-called main sequence). The
aim of their forward asteroseismic modeling is to fit their $n$ detected and
identified oscillation frequencies $f_i^{\rm obs}$ ($i=1,\dots,n$) with values
$f_i^{\rm th}$ predicted theoretically from stellar models and to evaluate in
this way the free parameters that are required as input to compute the stellar
interior properties adopted in these models.

A multitude of unknowns for the input physics occurs in stellar models of
  massive stars, even when fixing the microphysics \citep{Paper-I}.  Many of
  these unknowns in the macrophysics must be assigned constant values 
for computational reasons.
  \citet{Moravveji2015} determined a hierarchy of importance for asteroseismic
  modeling and showed that the two dominant uncalibrated ingredients in the
  models are core overshooting and envelope mixing.  Their relevant profiles,
  $D_{\rm ov}(r)$ and $D_{\rm mix}(r)$ (in the notation of Paper\,I), remain
  unknown. Because the bare minimum of stellar modeling in terms of input
  parameters is 4D without consideration of core overshooting and envelope
  mixing, asteroseismic modeling of massive stars is done by evaluating the {\it
    average level\/} of these two most important phenomena, rather than their
  shape. As such, $D_{\rm ov}(r)$ and $D_{\rm mix}(r)$ are simplified to two
  free parameters to estimate, denoted here as $f_{\rm ov}$ and $D_{\rm
    mix}$. This lies at the basis of asteroseismic model grids for intermediate-
  and high-mass star asteroseismology. Only after having achieved estimation of
  the average levels of $D_{\rm ov}(r)$ and $D_{\rm mix}(r)$ for sufficient
  pulsators can one move on to more sophisticated estimation of their profiles
  to assess theoretical concepts such as rotational mixing, wave transport, etc.
  \citet{Aerts2019} provides an extensive discussion of such phenomena and how
  asteroseismology is being used to tune them.  

In this work, we relied
on a 6D grid of stellar evolution models, covering a wide range in stellar mass $M$ from
2 to 20\,M$_\odot$. This grid was constructed in the framework of the European
FP\,7 SpaceInn Project.\footnote{This grid is publicly available from {\tt
    fys.kuleuven.be/ster/Software/grid-massive-stars}} It has been computed with
the open-source code MESA \citep[see the MESA instrument papers by][for
extensive documentation]{Paxton2011,Paxton2013,Paxton2015,Paxton2018}.  The grid
is based on the same input physics as used and described by
\citet{SchmidAerts2016}; the MESA inlist is provided in that paper and therefore
omitted here to avoid repetition.\footnote{Electronic versions of the MESA and
  GYRE inlists that lie at the definition of this grid can be retrieved from the
  MESA Marketplace: {\tt http://cococubed.asu.edu/mesa$_{-}$market/}} In brief,
the models are based on the following input physics: solar metal mixture by
\citet{Asplund2009}, OPAL opacity tables by \citet{Iglesias1996}, mixing-length
theory of convection with a fixed mixing-length parameter of 1.8 (expressed in
local pressure scale heights), and the Schwarzschild criterion for convective
stability.

This SpaceInn grid contains 352\,800 models of non-rotating stars
following the arguments of such an 
approach for asteroseismic modeling outlined in detail
  in \citet{Paper-I} and in \citet{Aerts2019}.  The grid covers
the evolutionary phase from the zero-age main sequence (ZAMS) until depletion of
central hydrogen at the terminal-age main sequence (TAMS), in steps of 0.01 in
central hydrogen mass fraction $X_c$.  Different chemical compositions in terms of
initial mass fractions were considered: initial hydrogen mass fractions $X$ of 0.68,
0.70, and 0.72, and initial metalicities $Z$ ranging from 0.012 to 0.020 in steps of
0.002.  The models were computed for exponential core overshooting in the
description by \citet{Herwig2000} and parametrized by a single parameter
$f_{\rm ov}$ (expressed in local pressure scale heights), ranging from 0.00 to
0.05, in steps of 0.01 (we refer to \citet{Paxton2011} for details of its
implementation in MESA).  
Given that all asteroseismic modeling results of intermediate-mass stars
based on {\it Kepler\/} data  demanded
``extra'' mixing in the radiative envelope to comply with the
detected trapping of gravity-mode oscillations in the near-core region,
in addition to core overshooting
\citep[e.g.,][]{Moravveji2015,Moravveji2016,SchmidAerts2016}, 
the
four levels of constant envelope
chemical mixing that were required so far for {\it Kepler\/} pulsators
were considered, i.e.,
$\log\,D_{\rm mix}=0, 1, 2, 3$\,cm$^2$\,s$^{-1}$. 
This implies that the SpaceInn grid is a 6D stellar model grid with 
parameters $(M, X, Z, X_c, f_{\rm ov}$, $D_{\rm mix})$. 
	
For chosen time steps along the evolutionary track, or equivalently $X_c$
values, the equilibrium models are perturbed in a linear framework to compute
oscillation modes and their accompanying frequencies. Here, we limit to
frequencies of coherent low-degree ($l$) zonal ($m=0$) gravity and pressure
modes, computed for each of the $\sim 70$ values of $X_c$ between the ZAMS and
the TAMS. These oscillation modes were computed with the public code GYRE
\citep{Townsend2013,Townsend2018}, adopting the Cowling and adiabatic
approximations as a good approach, as argued in \citet{Paper-I}. We
considered radial modes (degree $l=0$) and non-radial modes of degree $l=1, 2$,
for radial orders $n_{pg}\in [-50,+5]$, covering a range in frequencies (in an
inertial frame) of roughly $[0.05,50]\,$d$^{-1}$. In this way, we have about 62
million oscillation-mode frequencies available in the model grid.  

\citet{Johnston2018} and Mombarg et al.\ (in preparation) provide thorough
  discussions on how to treat parameter choices, their degeneracies, and their
  (often modest) deviation from linearity with respect to the stellar model
  properties. These two papers also discuss the way forward to deal with
  uncertainties following the statistical framework in \citet{Paper-I},
  building on earlier assessments of parameter correlations by
  \citet{Ausseloos2004}, \citet{Papics2014}, and \citet{Moravveji2015}. Manual
  forward modeling to estimate $f_{\rm ov}$ and $D_{\rm mix}$ so far ignored
  parameter correlations in the error estimation, also for low-mass stars with
  damped modes \citep{Deheuvels2016}.  As discussed at length in
  \citet{Johnston2018}, this is also the case in any binary modeling so far.
  The DNN provided here is a first step to improve upon manual analyses in terms
  of computational cost; it will need future upgrades to deal with correlations
  among the parameters and to deliver appropriate error estimation in a Bayesian
  framework, just as the manual analyses.

\section{Forward Asteroseismic Modeling of Coherent Oscillation 
Modes}

\citet{Paper-I} outlined in extensive
detail the principles of forward asteroseismic modeling of coherent oscillation
modes in the mass regime we consider here. In that work, the authors provided a
statistical methodological framework for such modeling, taking into account that
these oscillation modes depend non-linearly on the free parameters of stellar
models and that these parameters are correlated.  Their method is based on
Maximum Likelihood Estimation (MLE) and statistical model selection. We refer to
that paper for details, but repeat here that such modeling requires initial
screening of a large grid of stellar models in at least a 7D parameter space
when dealing with oscillations in rotating stars. For slow rotators, as we treat
here, one dimension can be avoided as one does not need to estimate the rotation
frequency of the star whenever it is so slow that rotation and the accompanying
Coriolis and centrifugal forces can be ignored in the pulsational computations.
In that case, one can reduce to the 6D parameter space as outlined for the
SpaceInn model grid in the previous section.

The method in Paper\,I requires the computation of various refined stellar model
grids, with different values for the input physics, in order to find a good
match between the observed oscillation frequencies and the theoretically
predicted frequencies, to within the measurement errors. The latter are
typically smaller than 0.001\,d$^{-1}$, such that the step sizes for
$(M, X, Z, X_c, f_{\rm ov}$, $D_{\rm mix})$ are required to be much smaller than
in the SpaceInn grid \citep[e.g.][for the first very detailed study and
application]{Moravveji2015}.  This procedure is only manageable (CPU-wise) to
apply to ensembles of stars when appropriate ``initial'' values for at least a
few of the six parameters can be achieved, such that the computation of
refined grids can be limited to appropriate sub-spaces of the
$(M, X, Z, X_c, f_{\rm ov}$, $D_{\rm mix})$ parameter space. Here, we develop a
deep learning method to achieve such an optimal initial guess, as an ideal
starting point to apply the methods in Paper\,I.

A major difference with the ML method developed by \citet{Bellinger2016}, which was  discussed above, is
that the observables we rely on as input for the machine learning only
constitute of asteroseismic quantities in the case of the intermediate- and high-mass stars
considered here.  Indeed, unlike for the low-mass sun-like stars, we do
not have high-precision spectroscopic estimates of the gravity $\log\,g$,
effective temperature $T_{\rm eff}$ and metalicity that can compete with the
asteroseismic data. For nominal {\it Kepler\/} data, the coherent modes have frequency errors at the level of
$\varepsilon_i/f_i^{\rm obs}<\!\!<\,0.1\%$, while the relative precisions for
$T_{\rm eff}$, $\log\,g$ and surface abundances from spectroscopy of stars in
the mass range considered here are typically above 10\%
\citep[e.g.][]{Morel2006,Morel2008,Martins2012a,Martins2012b,VanReeth2015}.  
So there is
at least two orders of magnitude difference in the quality of the spectroscopic and
asteroseismic information 
\citep[see Table\,1 in][for a detailed discussion on this matter]{Aerts2019}.  
Moreover, coherent modes in stars with a convective
core and a radiative envelope do not fulfil frequency separations as for high-frequency damped pressure modes. Hence, we
cannot rely on scaling relations to estimate the mass and radius as is the case
for solar-like oscillations.  In this respect, our set-up and application is
quite different from the case of low-mass stars with damped pressure modes as in, e.g.,
\citet{Appourchaux2012b,Gruberbauer2013,Corsaro2014,Davies2016,Bellinger2016,SilvaAguirre2017},
among others.

Whenever spectroscopic, interferometric or astrometric data at level of precision of the oscillation frequencies would become
available in the future,
they can of course be included in future asteroseismic modeling applications. However, our setup here only
relies on the oscillation frequencies and (partial) identification of the mode
degree $l$ and radial order $n_{pg}$ belonging to every measured oscillation
frequency $f_i^{\rm obs}$. The challenge is that we treat a mass range that is very large compared to applications done so far: $M\in [2,20]\,$M$_\odot$. We test in how far a deep neural network manages (or not) to pinpoint an appropriate narrow mass and age range from such a large spread in stellar parameters, from just a few (partially) identified oscillation mode frequencies without any other observational input.

\section{Deep Neural Networks}

Neural networks, and deep neural networks (DNNs) in particular, have received a lot of attention recently, mainly because of the advancements
in computer vision and natural language processing. Although the theory behind
neural networks is quite old, only now are the datasets sufficiently large and
the required processing power good enough to attain high accuracies on many
different types of computational problems. In this work, a DNN is trained on an
asteroseismological grid of intermediate-mass and high-mass stars to predict the
frequencies of their coherent oscillation modes. This is a typical
  regression problem. 

\begin{figure}
\centering
  \includegraphics[width=.45\textwidth]{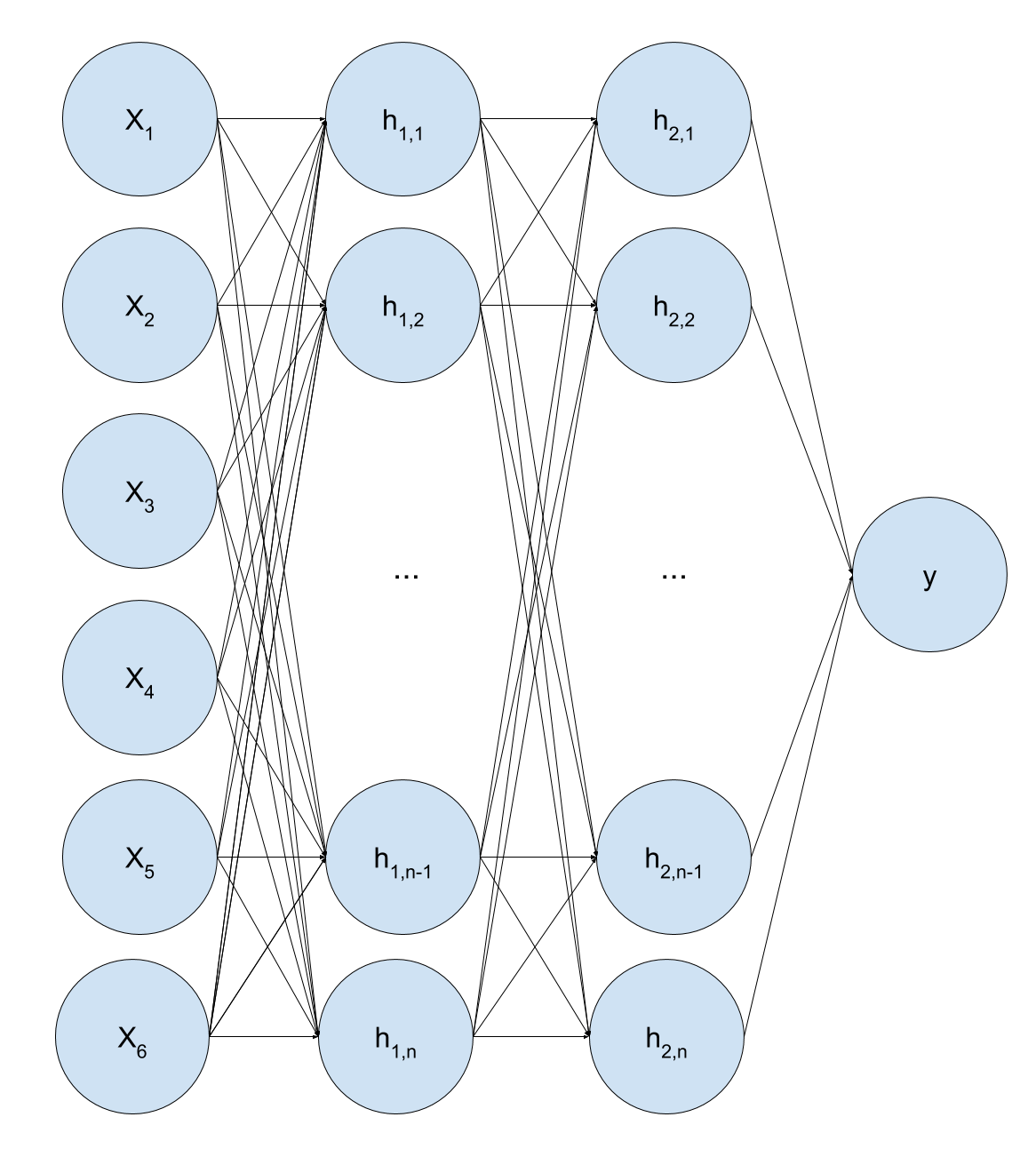}
\caption{Schematic view of a dense network. In a dense network, every neuron is connected with every neuron in the previous layer.}
\label{fig:dense-network}
\end{figure}

The neural network architecture used in this work is a fully connected or dense
network. A schematic view of a dense network is shown in Fig.\,\ref{fig:dense-network}.
The input vector in this example is 6-dimensional and the output is
1-dimensional. In between are two layers that are $n$-dimensional. The layers have
6, $n$, $n$ and 1 neurons, respectively, in this case. The value of each neuron, except
the input neurons, is determined by the following function:

\begin{equation}
y=f\left(\sum_j w_j x_j  + b\right),
\end{equation}

\noindent where $f\left(\cdot\right)$ is called the activation function, $w_j$ is a weight
parameter between two neurons (the arrows in Fig.\,\ref{fig:dense-network}) and
$b$ is a bias term. In order to add non-linearities to the neural network, the
activation function must be non-linear. A typical choice is the Rectified Linear
Unit \citep[ReLU,][]{icml2010_NairH10}. This function is defined as $f(x)=x$ if
$x>0$ and $f(x)=0$ if $x\leq 0$.

By starting from the input layer and calculating all values of all neurons
layer-by-layer, the output is finally computed.  Initially, the weights and
biases are set to random numbers and during training the weights are adjusted to
correctly predict a predetermined training set. To validate that the
  network has correctly learned to predict the occurring features, the network
  is evaluated on a separate validation set. Typically, the validation set is
  10\% to 20\% of the total dataset.  For regression problems, this typically
is a mean-squared error between the prediction by the neural network and the
actual output belonging to a particular training sample. By backpropagating the
error through the network, the weights and biases are updated to improve the
cost function in the next iteration \citep{Chauvin:1995:BTA:201784}.

A typical caveat of training neural networks is overfitting. This means the
neural network learns to replicate the exact results of the training set,
instead of generalizing over the particular samples and learning the underlying
correlations. This can be solved by adding a penalty term for high-valued
weights, which is called regularization \citep{Ng:2004:FSL:1015330.1015435}. One can verify if a network is overfitted by comparing the training and validation set accuracy. If the training loss is decreasing but the validation loss is increasing, the network is overfitting. A
more thorough introduction into deep learning and the validation approach we
followed here is given in the Appendix of
\citet{Caron:2017udl}, to which we refer for further details.

\section{Deep neural networks applied to asteroseismic modeling}
\label{sec:results}

The aim of this work is to train a neural network that is able to predict the
frequencies of the coherent pulsation modes of a given star with detected
oscillation frequencies belonging to identified modes $(l,n_{pg})$. We want to
achieve the capacity to quickly search the parameter space spanned by the
stellar parameters to find the optimal values for given measurements of the
oscillations, without relying on any other information.

The choice of using deep neural networks for this problem has several
reasons. Firstly, the size of the dataset is quite large. From the grid, around
62 million oscillation modes are extracted and the objective is to predict the
frequency belonging to specific modes given by a wide range of stellar
parameters. The aim is not to train on parameters of a particular star, but for
all possible stellar parameters that are available in the dataset in one
go. This forces the machine learning algorithm to generalize over all parameters
and learn the underlying structure. It also means that the network only has to
be trained once, and then can be applied to all stars that reside somewhere
inside the training grid. Although training can take some time (around 6 hours
for this work), evaluation is typically very fast (for this work in the order of
microseconds per inference) and the network size is not very large (the
compressed network file of this work is 370MB).  Furthermore, by adding a batch
normalization layer as the first layer inside the network, the issue of
incorrectly weighting parameters as stated in \cite{Bellinger2016} is no longer
present. The normalizing layer normalizes all data before it is fed to the
trainable parameters.

In \cite{Bellinger2016} it is also stated that deep learning based regression is
unconstrained. In this work that is indeed the case. However if the network
accuracy is high enough this does not have to be a problem. If it would be a
problem, it is trivial to design a constrained activation function that is
bounded between two values. For example the hyperbolic tangent activation
function is bounded between $\pm 1$ and can trivially be scaled and translated
to get a bound between two values.
A downside of deep learning compared to random forests is that it is fairly easy
with random forests to get insights into why particular predictions are being
made, while this information is not accessible from a deep neural
network. Training a deep network is also typically more difficult due to the
high number of hyperparameters that need to be set correctly to
optimally train on the data.

\begin{table}
\centering
\caption{The neural network architecture used to train and predict
the oscillation  frequencies of coherent low-degree modes. 
Every fully connected layer has a ReLU activation function, 
unless otherwise specified between brackets. Every fully
  connected layer has L2 regularization to prevent overfitting.}
\label{tab:nn-arch}
\begin{tabular}{lll}
Layer \# & Layer type          & Number of neurons \\
1        & Input layer         & 8                 \\
2        & Batch normalization &                   \\
3        & Fully connected     & 500               \\
4        & Fully connected     & 500               \\
5        & Fully connected     & 500               \\
6        & Fully connected     & 500               \\
7        & Fully connected     & 500               \\
8        & Fully connected     & 500               \\
9        & Fully connected     & 5000              \\
10        & Fully connected     & 5000              \\
11        & Fully connected (tanh)     & 5000              \\
12        & Fully connected (linear)     & 1                
\end{tabular}
\end{table}

In this work, a DNN is trained using TensorFlow
  \citep{tensorflow2015-whitepaper} and TFLearn \citep{tflearn2016} on an
  asteroseismic grid of intermediate- and high-mass stars to predict
  the frequencies of their coherent oscillation modes. The neural network
architecture is shown in Table\,\ref{tab:nn-arch}, which has a total number of
53,777,033 trainable parameters. Different neural network architectures
  have been evaluated and the architecture described in Table\,\ref{tab:nn-arch}
  provided the best accuracy. In addition, different hyperparameter combinations
  have been evaluated. The network is trained using the Adam optimizer
  \citep{DBLP:journals/corr/KingmaB14} with a learning rate of $10^{-3}$ and a
  batch size of 2048.  

The input data for the neural network is 8D: the six
stellar parameters ($M$, $X$, $Z$, $X_c$, $f_{\rm ov}$, and $D_{\rm mix}$) and
two integer mode quantum numbers connected with the frequency of the zonal mode:
$l$ and $n_{pg}$. The output of the network is the mode frequency $f^{\rm th}_i$
for all modes $i$ with degree 0, 1, or 2 and for all radial orders $n_{pg}\in
[-50,+5]$ (cf.\ Fig.\,\ref{fig:dense-network}).

A typical asteroseismic measurement consists of a list of frequencies
  with corresponding $n_{pg}$ and $l$ values. Often, the $n_{pg}$ and $l$-values 
are not (all)
  known, but can be one of several values or depend on the values for (one of)
  the other frequencies. For example $n_{pg;i+1} < n_{pg;i}$, where $n_{pg;i}$ is
  the $i$-th $n_{pg}$ value corresponding to a particular frequency. As the
  neural network predicts one frequency from a list of stellar parameters and a
  combination of $n_{pg}$ and $l$, another method is needed to find the optimal
  combination of $n_{pg}$ and $l$ values, given a measurement of multiple
  frequencies. The method used in this work is a genetic algorithm that tries to
  find the best fit values for a given list of observed frequencies. For every
combination of the six stellar parameters, all frequencies are calculated for
$l \in \{0,1,2\}$ and $n_{pg}$ between $-50$ and $+5$ in one batch. On an Nvidia
GTX1080, these 168 evaluations take about 6 milliseconds. Then, the best fitting
combination of $n_{pg}$ and $l$ values are chosen following a cost function
specified below. The cost is minimized using a particle filter, which follows
the same procedure as in \citet{Caron:2015wda}:
\begin{enumerate}
\item Start with a random sample of $N$ points;
\item Take the $n$ points with the least error (lowest cost);
\item Sample $m$ points around the $n$ points taken from the previous step,
  using a Gaussian prior with as mean the parameter values and standard deviation $\sigma_1$ times the mean;
\item Repeat steps 2 and 3 $I$ times;
\item Repeat steps 2, 3 and 4 also with a standard deviation of $\sigma_2$ and $\sigma_3$ times the mean.
\end{enumerate}

The cost function mentioned above is different than the cost function
  of the neural network. In this step, the neural network is already trained and
  not optimized anymore. In addition, the loss of the neural network is not
  propagated to the cost function of the particle filter. This is because the
  error of the neural network is constant over all predictions. In a future
  update the neural network output will also contain uncertainty information
  (see Section\,\ref{sec:conclusion}). The cost function that is to be minimized
  is one of the following options, depending on how much is known about the
  measured oscillation frequencies: 

\begin{enumerate}
\item If a list of frequencies with exactly known $l$ and $n_{pg}$ values is available,
  those values are compared to the predicted values. The cost of the predicted
  point with respect to the measurement is $\chi^2=\frac{1}{N} \sum_i^n
  \left(f_i^{\rm th}-f_i^{\rm obs}\right)^2$, where $f^{\rm th}$
represents predicted frequencies and $f^{\rm obs}$
measured frequencies, following the notation of Paper\,I.

\item If for a list of frequencies $n_{pg}$ is available, but $l$ can be one of
  multiple values, all possible $l$ values are tested and the minimum error is
  used in the same cost function as the previous item.

\item If for a list of frequencies only is known that the frequencies are
  decreasing for decreasing $n_{pg}$, the best fitting $n_{pg}$ value is used in
  the list of possible $n_{pg}$ values, bounded by $(-50, 5)$ or different
  bounds if known.

\item If for a list of frequencies only their relative differences are known,
  the $\chi^2$ as defined in the first item is calculated for every possible
  combination of $n_{pg}$ and $l$ values within the bounds $(-50, 5)$ or
  different bounds if known.
\end{enumerate}

The analysis setup above defines a recipe to find the best fit star parameters
in a straightforward way:
\begin{enumerate}
\item Given a measurement, define the cost function set up given by the above
  possibilities;
\item Using this cost function, find the best stellar parameters using a particle
  filter that evaluates on the network output.
\end{enumerate}

It is important to note that the particle filter could also be run on
  the actual simulation instead of the neural network in order to find the
  optimal combination of stellar parameters for a given list of
  frequencies. However, given the required amount of iterations to minimize the
  cost function this would be too time-consuming. Once the particle filter is
  used to find stellar parameters, the trained neural network is fixed and
  assumed to give correct output. This means the neural network is merely used
  as a fast replacement for the underlying physics simulator. The particle
  filter has seven hyperparameters. The choice of these determine how long the
  particle filter runs and how well it is able to find the global minimum. The
  choices of the hyperparameters are listed in 
Table\,\ref{tab:pf-hyperparameters}. To
  illustrate the effect of different hyperparameters, two choices have been made
  and the results for both are in each case overplotted in the application
  figures below as gray and the blue points. The total amount
  of neural network inferences is $N + 3 \cdot i \cdot n$. The goal of the
  particle filter is to find the global minimum in as few iterations as
  possible. The hyperparameter choices in 
Table\,\ref{tab:pf-hyperparameters} are
  found to be the most optimal combination to find the global minimum while
  keeping the total amount of neural network inference as low as possible. The
  blue points can be seen as a quick scan, where one finds the region of interest
  for the six stellar parameters. The gray points zoom in on the optimal
  combination more intensely. 
These values were found by incrementally increasing the
  hyperparameter values and evaluating the accuracy while keeping in mind that
  the number of evaluations should be as low as possible.

\begin{table}[h!]
\centering
\caption{Choices of hyperparameters for the particle filter algorithm. 
Two choices have been made to illustrate the effect for different 
hyperparameters: gray and blue. The names of these combinations 
resemble the point color in the resulting plots of the applications to
stars. See text for explanation.}
\label{tab:pf-hyperparameters}
\begin{tabular}{lll}
Parameter      & Gray & Blue     \\
\hline
$N$ & 5000  & 1000         \\
$n$ & 2000  & 100         \\
$m$ & 200   & 10         \\
$i$ & 10   & 10         \\
$\sigma_1$ & 0.4   & 0.4         \\
$\sigma_2$ & 0.1   & 0.1         \\
$\sigma_3$ & 0.01   & 0.01         \\
\hline
\end{tabular}
\end{table}

\begin{figure}[h!]
\centering
\includegraphics[width=0.8\textwidth]{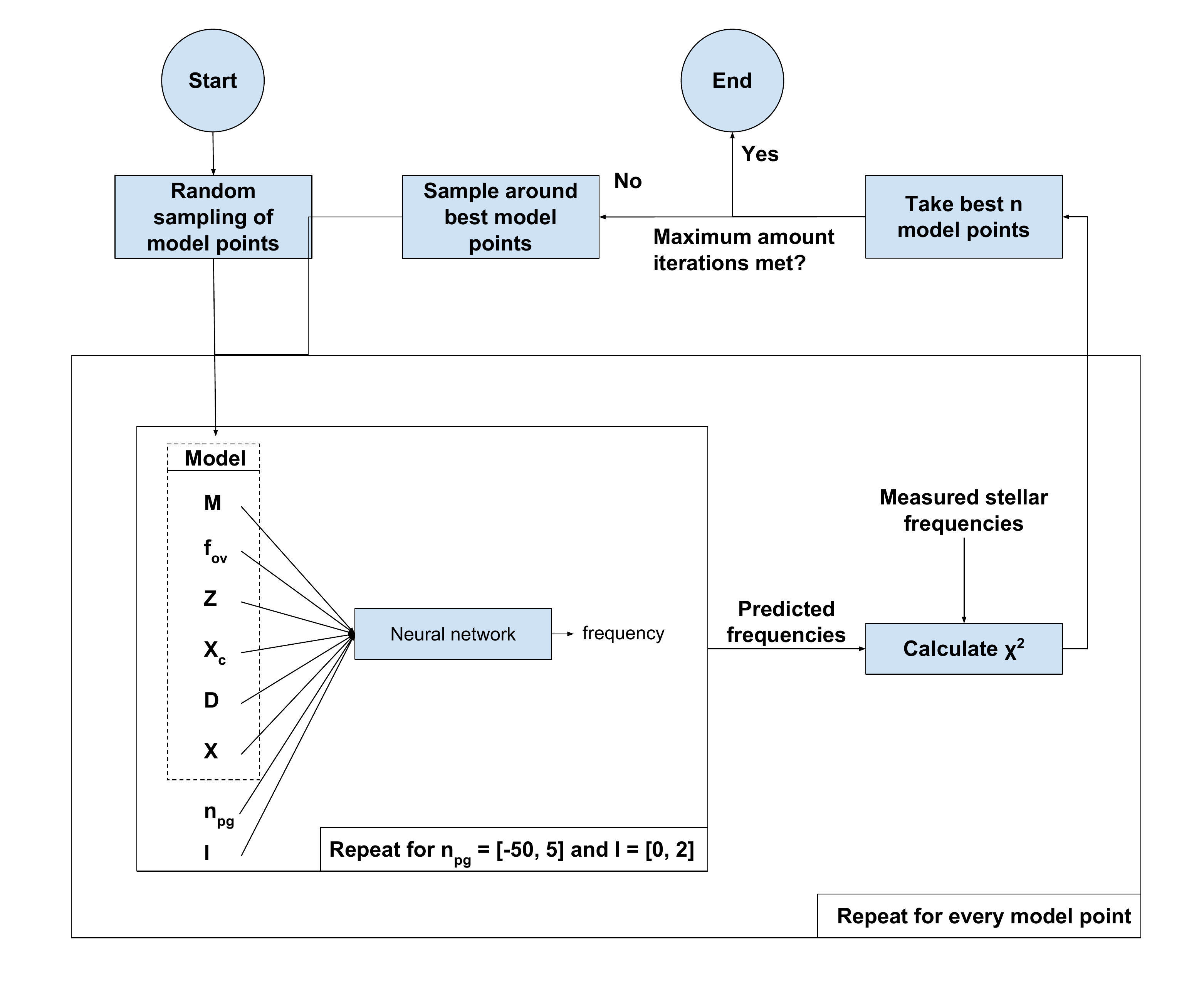}
\caption{Visualization of the whole pipeline to find the best fitting stellar
    parameters for a given set of measured frequencies. The method is a
    maximum-likelihood evaluation over a DNN, which is trained to replicate a
    physics simulator. The result is a prediction of model parameters that best
    describe the data.}
\label{fig:pipeline}
\end{figure}

A visualization of the
  full pipeline is shown in Fig.\,\ref{fig:pipeline}.
  This method is able to find the minimum of a given set of measurements
in about 30 seconds using an Nvidia GTX1080 GPU. Note that this excludes
  the training time of the actual neural network. However, the neural network
  only has to be trained once and then is applicable to all stars within the
  grid of the training dataset. The whole pipeline of finding the optimal set of
  stellar parameters using the particle filter and DNN is abbreviated below as the
  PF-DNN method.  

\newpage

\section{Results}

\begin{figure}[h!]
\centering
\includegraphics[width=0.8\textwidth]{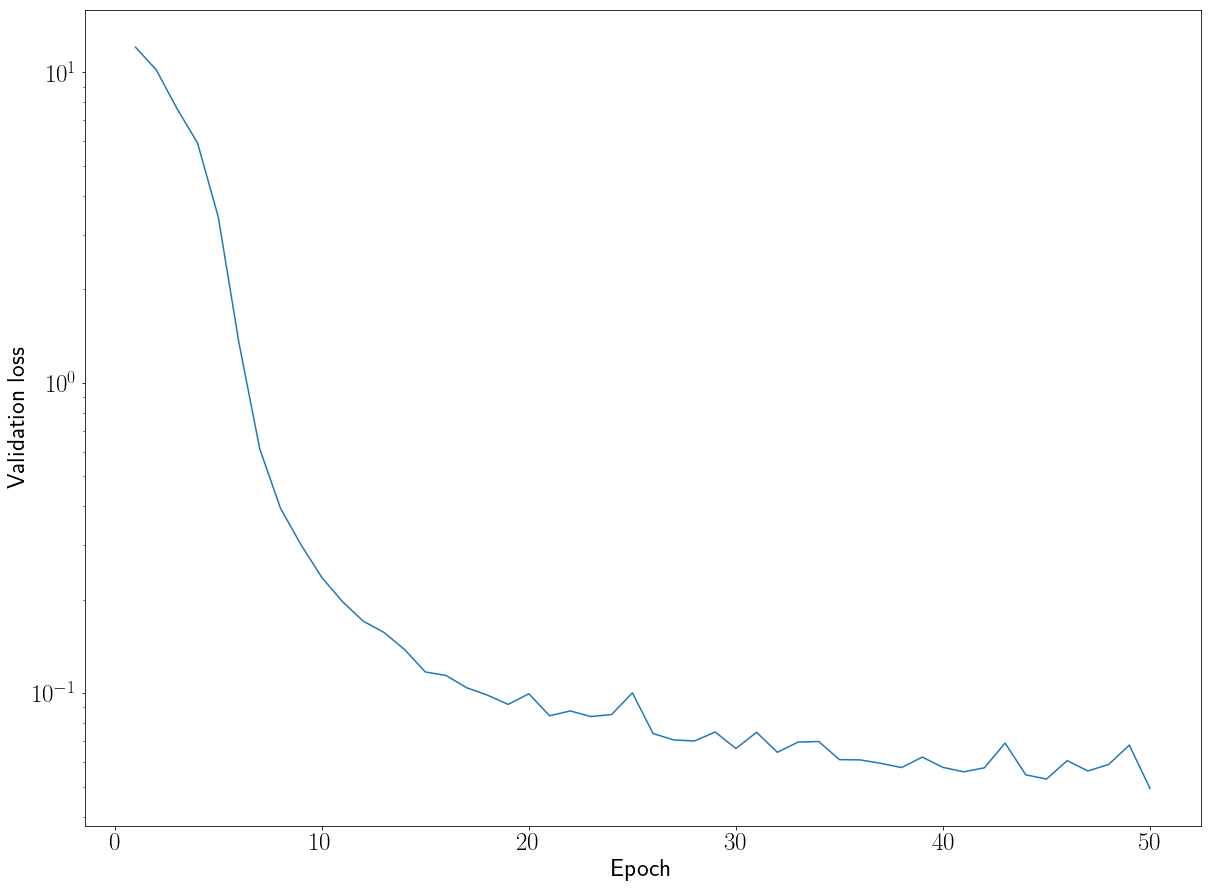}
\caption{Validation loss per epoch of the neural network.}
\label{fig:loss}
\end{figure}

\begin{figure}[h!]
\centering
\includegraphics[width=0.8\textwidth]{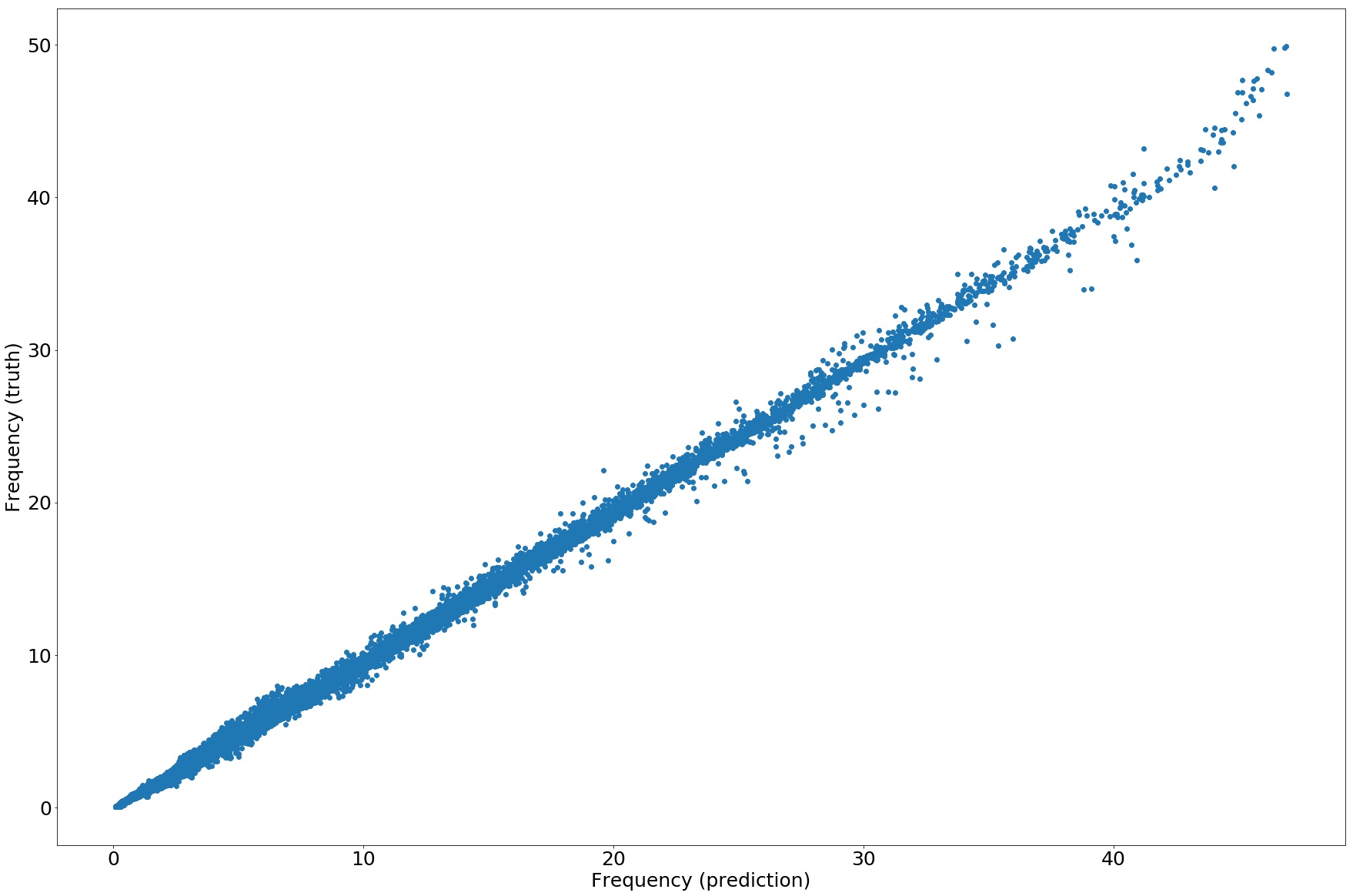}
\caption{Visualization of the performance of the network. Plotted are the
  network predictions and the true values (from the dataset), for all stellar
  parameters and frequency modes. In this plot, 100,000 points from the
  validation data are shown.}
\label{fig:acc}
\end{figure}

In Fig.\,\ref{fig:loss} the loss on the validation set is plotted for
  50 epochs. The mean-squared-error loss of the final result is 0.05. On an
  Nvidia GTX 1080 it takes around six hours to train the network. 
Figure\,\ref{fig:acc} shows the predictive power of the neural network. As
  can be seen in the figure, the neural network almost always predicts the same
  frequency as is in the grid. Only for high frequencies the neural network
  underpredicts the frequency. This is probably due to an imbalance of the
  training data with respect to the frequency: in the plot are 100,000 random
  points from the validation set. As can be clearly seen, the distribution of
  frequencies is not flat. This could probably be improved by using for example
  pool based active learning, which will be explored in the next version of the
  pipeline.   

As a proof-of-concept, the procedure defined in the previous sections is applied
to seven selected slowly rotating stars of spectral type B for which forward
asteroseismic modeling has been achieved in the past. We stress that our DNN
is a preprocessing method, meant to provide a meaningful 
sub-space within a
huge 6D grid of core hydrogen burning models with extensive coverage in mass,
based on the input of a few (identified or not) coherent oscillation modes. If
this works out, then we found a very fast way to define and start the
computation of detailed and refined multi-D grids of stellar models for forward
modeling, following the methodology of Paper\,I.  Our approach and results have
to be evaluated with this aim in mind.

We focus on five $\beta\,$Cep stars with low-order pressure and gravity modes
and two slowly pulsating B stars (SPBs) with high-order gravity modes. All seven
pulsators are very slow rotators, such that we can limit to zonal modes ($m=0$).
While the past CoRoT and {\it Kepler\/} and current BRITE space missions
delivered good asteroseismic data for several intermediate-mass SPBs
\citep[e.g.,][]{Papics2017,Kallinger2017,Szewczuk2018}, this was not the case
for the more massive $\beta\,$Cep stars. However, the TESS mission
\citep[][launched successfully on 18 April 2018]{Ricker2016} will deliver data
of hundreds of such pulsators very soon.

An important remark is that the five $\beta\,$Cep stars have been modeled
asteroseismically with stellar evolution and pulsation codes that are completely
independent from those that were used to construct the SpaceInn grid. They
therefore constitute by far the most important test for our PF-DNN
  method, because the used model grid and the DNN methodology are both new and
are built upon independent ingredients from the previous decade-old
applications. The two SPBs were modeled with the MESA and GYRE codes, which also
lie at the basis of the SpaceInn grid. For those, the test is therefore mainly
at the level of the PF-DNN strategy,  and not so much on the input
physics of the model grid. Our major aim is to test if the PF-DNN
  method can derive appropriate ranges for the mass and evolutionary stage of
the stars, keeping in mind that degeneracies occur between $(M,Z,f_{\rm ov})$.

\subsection{The $\beta\,$Cep star HD\,129929}

Apart from the Sun, HD\,129929 was the first star in the core hydrogen burning
phase for which an asteroseismic estimate of the rotation rate in the stellar
interior could be obtained \citep{Aerts2003}. This was achieved from 21 years of
single-site multi-color photometric monitoring at mmag precision, which resulted
in the detection of a radial mode, a dipole triplet, and two members of a
quadrupole quintuplet \citep{Aerts2004}.  A spectroscopic estimate of $T_{\rm eff}$
and $\log\,g$ added to the photometry led \citet{Dupret2004} to a near-core
rotation frequency of 0.013\,d$^{-1}$ and a $\sim$4 times faster near-core than
envelope rotation, in agreement with the spectroscopically measured
$v\sin\,i$. The authors also identified the need for non-zero yet low core
overshooting.  This phenomenon was modeled as convective
penetration and required a step-overshoot value of
$\alpha_{\rm ov}=0.10\pm 0.05$  (expressed in local pressure scale heights). Aside from the core overshooting, the star's
mass $M$, metalicity ($Z$), and age (i.e., through $X_c$) were estimated (cf.\
Table\,\ref{tab:star1}). This asteroseismic forward modeling was done
``manually'' at a time when CPU power was much more limited than currently, by ``meandering'' in a sparse grid. Finding an appropriate asteroseismic model at that time required typically weeks/months to achieve, depending on the experience of the user.

\begin{figure}[h!]
\centering
\includegraphics[width=\textwidth]{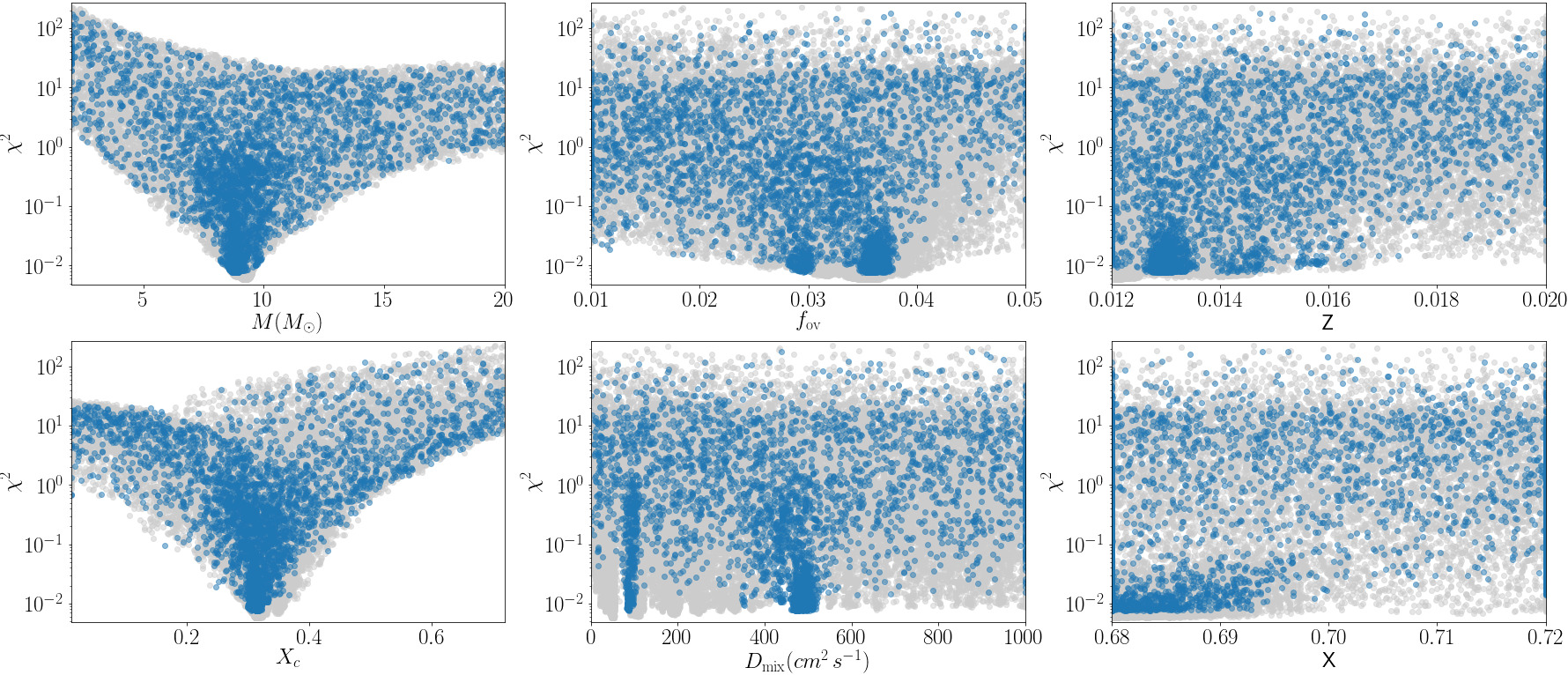}
\includegraphics[width=\textwidth]{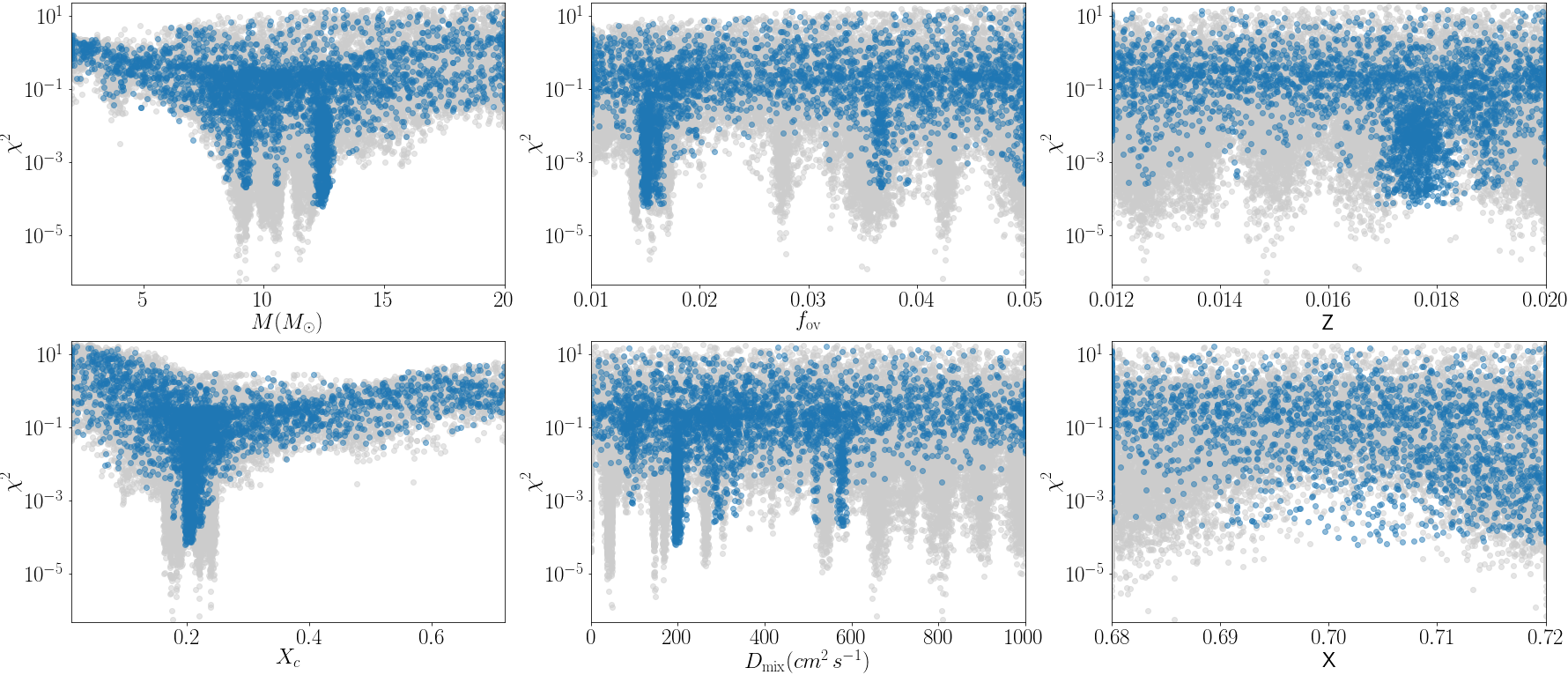}
\caption{PF-DNN method applied to HD\,129929, based on its three identified low-order 
modes. Upper 6 panels: by providing both the three mode degrees $l$ and the radial orders $n_{pg}$ to the PF-DNN method; lower 6 panels: by only providing the three mode degrees $l$.}
\label{fig:star1}
\end{figure}

\begin{table}[h!]
\centering
\caption{Best PF-DNN fit points (resp.\ blue (sparser) and gray (densest) dots in Fig.\,\ref{fig:star1}) for HD\,129929 compared with the asteroseismic solution
  in the literature.}
\label{tab:star1}
\begin{tabular}{llll}
Parameter      & PF-DNN Value (gray) & PF-DNN Value (blue)        & \citet{Dupret2004} \\
\hline
$M$ (M$_\odot$)                    & 9.23  & 8.81       &   $\sim$9.35  \\
$f_{\rm ov}$                 & 0.0365      & 0.0356         & $\alpha_{\rm ov}=0.10\pm 0.05$ \\
$Z$                          & 0.012       & 0.0130          & $\in [0.016,0.022]$\\
$X_c$                        & 0.314       & 0.314          & 0.353 \\
$D_{\rm mix} (cm^2\,s^{-1})$  & 48         & 475  & --       \\
$X$                          & 0.680       & 0.68          & 0.7 (fixed) \\
\hline
\end{tabular}
\end{table}

Here, we only use the three identified zonal ($m=0$) modes of HD\,129929
(ignoring the spectroscopic information) as a first test case for our PF-DNN method. It
concerns the observed frequencies 6.590940\,d$^{-1}$ with $l=0, n_{pg}=+1$,
6.978305\,d$^{-1}$ with $l=1, n_{pg}=+1$, and 6.461699\,d$^{-1}$ with
$l=2, n_{pg}=-1$ \citep{Aerts2003,Aerts2004}. The outcome is shown in the upper six panels of 
Fig.\,\ref{fig:star1} and the
parameters identified as ``best'' model by the trained network are presented in
Table\,\ref{tab:star1}. First of all, one should not restrict to the bare minimum given by the PF-DNN method as listed in Table\,\ref{tab:star1} but always consider the morphology of the whole PF-DNN outcome as in 
Fig.\,\ref{fig:star1}, as broad minima may occur.
Moreover, we point out that the PF-DNN uses diffusive convective
overshooting with the radiative temperature gradient in the overshoot zone
rather than convective penetration with the adiabatic gradient.  The relation
between $\alpha_{\rm ov}$ and $f_{\rm ov}$ is about a factor 10 to 12
\citep{Moravveji2015,ClaretTorres2017} and must be taken into account when evaluating the PF-DNN
capacity. Keeping this in mind, we deduce from Fig.\,\ref{fig:star1} and
Table\,\ref{tab:star1} that the PF-DNN does a remarkably good job when considering the dense network (indicated as gray solution in Table\,\ref{tab:star1}), particularly
keeping in mind that the metalicity cannot be well constrained from forward
modeling of high-mass stars and that $M$ and $f_{\rm ov}$ are correlated in the case of B-type pulsators
\citep[e.g.,][]{Walczak2013,Moravveji2015,Pedersen2018}.
It can also be deduced 
from the six upper panels in Fig.\,\ref{fig:star1}  that the initial hydrogen content $X$ and the level of
chemical mixing in the envelope $D_{\rm mix}$ are not well constrained. This is generally the case for the $\beta\,$Cep stars treated here, and in particular for HD\,129929, 
given that there are only three detected oscillation modes to estimate six parameters. The same limitation
occurred from the manual forward modeling done 15 years ago.

For most applications, one does not know the radial order of the modes connected with the measured frequencies. Any first screening of models must then be based on the provision of the mode degrees $l$, without providing the radial orders $n_{pg}$. We used the case of HD\,129929 to test how the lack of values for $n_{pg}$ affects the results. The outcome can be seen in the lower six panels of Fig.\,\ref{fig:star1} and 
reveals that, without information on the radial order, one finds several sharp minima for the mass and evolutionary stage. The secondary minimum corresponds to a more massive and more evolved model for which the radial mode is the first overtone rather than the fundamental mode. This model has a somewhat lower convective core overshooting and is older. This exercise shows that, whenever the radial order of the modes is unknown quantitatively but only in terms of relative ranking among the detected frequencies, one can still get a good solution, but it is to be advised that the  several options for the mass are considered for refined and detailed modeling, e.g.\ with the methods in Paper\,I.

\subsection{The $\beta\,$Cep star HD\,157056}

In terms of structure in the frequency spectrum, the $\beta\,$Cep star HD\,157056 is almost a copy of HD\,129929 \citep{Briquet2007}, except that its detected rotational splitting is in agreement with rigid rotation with rotation frequency 0.107\,d$^{-1}$ throughout the star. This is eight times faster than the rotation of HD\,129929, but still very slow compared to the periods of order hours of its low-order oscillation modes and far below its critical rotation rate. It is therefore still fine to ignore the effect of the Coriolis force on the oscillation frequencies, particularly for the zonal modes.

\begin{figure}[h!]
\centering
  \includegraphics[width=\textwidth]{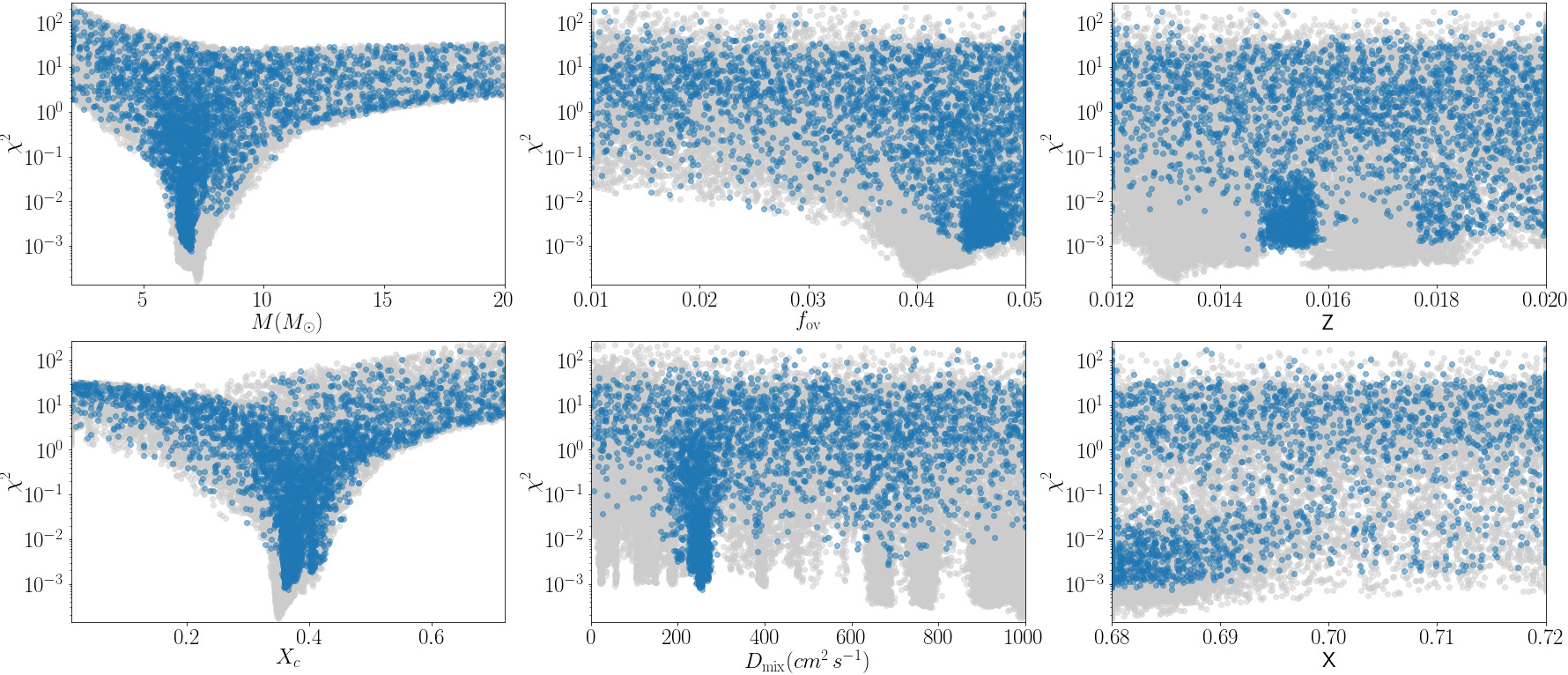}
\caption{PF-DNN method applied to HD\,157056, based on its three identified low-order 
modes.}
\label{fig:star2}
\end{figure}

\begin{table}[h!]
\centering
\caption{Best PF-DNN fit point for HD\,157056 
compared with the asteroseismic solution  in the literature.}
\label{tab:star2}
\begin{tabular}{llll}
Parameter                   & PF-DNN Value (gray) & PF-DNN Value (blue)    & \citet{Briquet2007}\\
\hline
$M$ (M$_\odot$)                   & 7.21  & 6.98        & 8.2$\pm$0.3    \\
$f_{\rm ov}$                 & 0.0402     & 0.0448    &    $\alpha_{\rm ov}=0.44\pm 0.07$      \\
$Z$                          & 0.0132     & 0.0151     & $\in [0.009, 0.015]$       \\
$X_c$                        & 0.349      & 0.365     & $0.38\pm 0.02$     \\
$D_{\rm mix} (cm^2\,s^{-1})$  & 988       & 256   & --       \\
$X$                          & 0.680      & 0.686           & $\in [0.71, 0.7211]$ \\
\hline
\end{tabular}
\end{table}

The zonal oscillation-mode frequency values of HD\,157056, i.e., 7.4677\,d$^{-1}$ for $l=0, n_{pg}=+1$, 7.8742\,d$^{-1}$ for $l=1, n_{pg}=+1$, and 7.204\,d$^{-1}$ for
$l=2, n_{pg}=-1$, are higher than the corresponding ones of HD\,129929. Since it concerns zonal pressure modes, which are connected with standing acoustic waves, this implies that the star must be smaller and hence less massive than HD\,129929. This was confirmed by the asteroseismic modeling results obtained by \citet{Briquet2007} listed in Table\,\ref{tab:star2}. A striking result found by these authors is the large value of $\alpha_{\rm ov}=0.44\pm 0.07$ for the core overshooting, where their models were based on convective penetration.

\citet{Briquet2007} also investigated models with atomic diffusion but in a description that ignores radiative levitation.  It is well known that atomic diffusion in models of stars with a radiative envelope implies the helium and metals to sink too strongly, leading to incorrect surface abundances. One usually counter-balances this too strong settling by adding an amount of turbulent mixing in the superficial envelope layers. \citet{Briquet2007} showed that the inclusion of atomic diffusion along with such artificial turbulence, at a level that brings the predicted surface abundances in agreement with the observed values, does not change the asteroseismic parameter estimates appreciably.

Our PF-DNN results of HD\,157056 are shown in Fig.\,\ref{fig:star2}. In this case, it can be seen that a deep minimum and a secondary minimum for the mass occur for the blue dots, while one cleaner minimum occurs for the gray dots. The most important conclusion, however, is independent of which version of the DNN we adopt: 
as reported in 
Table\,\ref{tab:star2}, the DNN confirms the high level of core overshooting for this star. Further, the DNN results in a different $(X,Z)$ combination and a 1\,M$_\odot$ lower mass compared to the forward modeling. It is well known that a mass-metalicity relation occurs when fitting low-order pressure modes in $\beta\,$Cep stars \citep[lower mass is accompanied by a higher metalicity: Fig.\,2 in][]{Ausseloos2004}. Moreover, as already mentioned, the core overshooting and mass are also correlated. Taking this into account, the PF-DNN solution is appropriate. In particular, the PF-DNN manages to pinpoint the correct evolutionary stage of the star.

\subsection{The $\beta\,$Cep star HD\,886}

The $\beta\,$Cep star HD\,886 was long known to be a slowly rotating monoperiodic radial pulsator, but space photometry assembled with the MOST satellite revealed thirteen non-radial oscillation modes aside from the dominant radial mode. Among these thirteen frequencies are seven low-order modes and six gravity modes
\citep{Handler2009}. While forward modeling has not been done for this star yet, \cite{Walczak2013} did make an extensive comparison of the observed oscillations and those predicted by theoretical models, with the main aim to test the mode excitation theory. They managed to identify five of the fourteen detected modes,
assuming zonal modes: the three low-order modes with 6.58974\,d$^{-1}$ for $l=0, n_{pg}=+1$, 6.01616\,d$^{-1}$ for $l=1, n_{pg}=-1$, 9.1092\,d$^{-1}$ for $l=1, n_{pg}=+2$, and the two high-order gravity modes with 0.8352\,d$^{-1}$ for $l=1, n_{pg}=-9$, 0.63551\,d$^{-1}$ for $l=1, n_{pg}=-12$. The parameters of
their ``Model 2'' are listed in Table\,\ref{tab:star3}. For this star, the shift induced by rotation is less than 0.005\,d$^{-1}$ \citep{Walczak2013} so that rotation can be ignored in the application. 

\begin{figure}[h!]
\centering
\includegraphics[width=\textwidth]{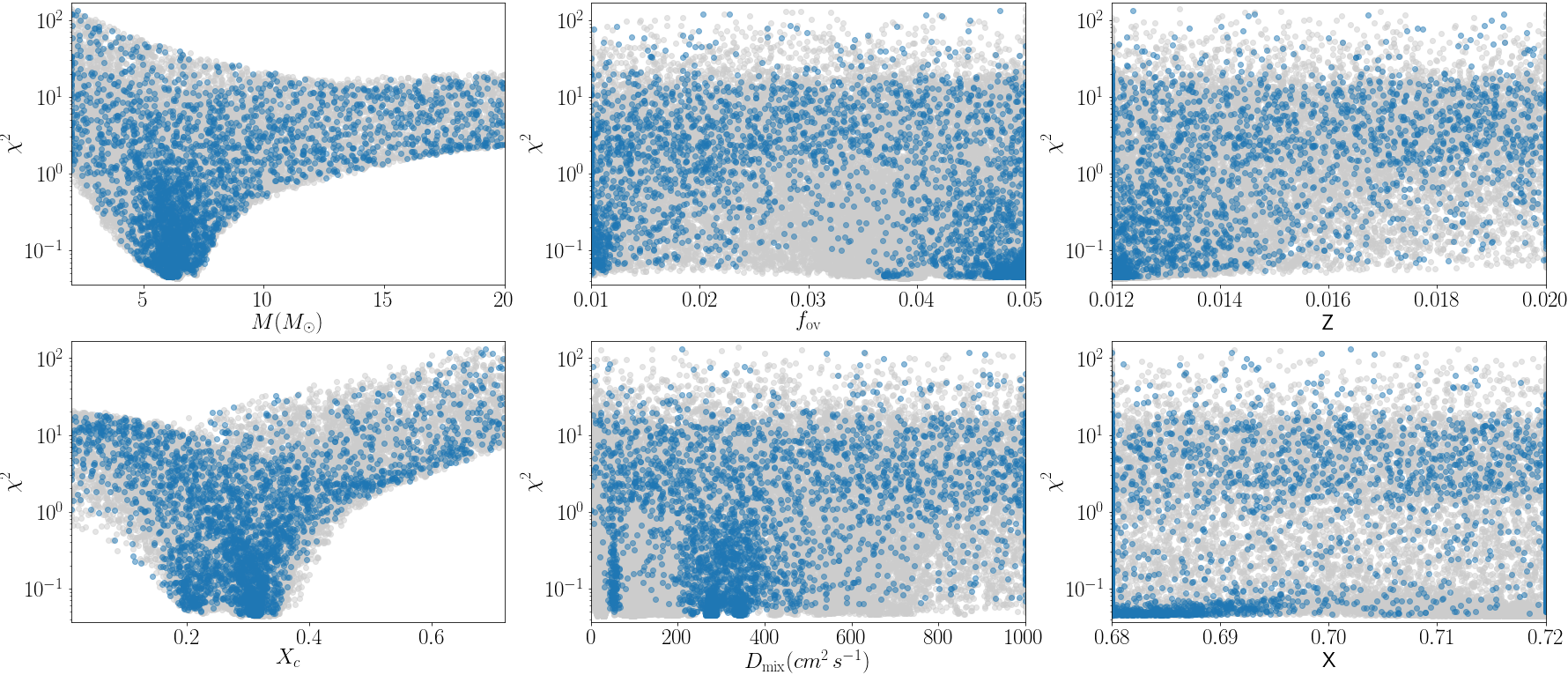}
\caption{PF-DNN method applied to HD\,886, based on its three low-order 
modes and two high-order gravity modes.}
\label{fig:star3}
\end{figure}

\begin{table}[h!]
\centering
\caption{Best PF-DNN fit point for HD\,886 
compared with the asteroseismic solution  in the literature.}
\label{tab:star3}
\begin{tabular}{llll}
Parameter                    & PF-DNN Value (gray) & PF-DNN Value (blue)         & \citet{Walczak2013} \\
\hline
$M$ (M$_\odot$)              & 6.43     &    6.13    & 7.9   \\
$f_{\rm ov}$                & 0.0388    &    0.0499 &  $\alpha_{\rm ov}\simeq 0.25$     \\
$Z$                          & 0.012    &    0.012         & 0.0135    \\
$X_c$                        & 0.312    &    0.312           &   -- \\
$D_{\rm mix} (cm^2\,s^{-1})$  & 269     &    269    & --        \\
$X$                          & 0.68     &    0.68            & 0.71 (fixed) \\
\hline
\end{tabular}
\end{table}

Our PF-DNN results of HD\,886 are shown in Fig.\,\ref{fig:star3} and the two minima of the network are listed in Table\,\ref{tab:star3}. Again, we find a lower mass compared with the interpretation of the frequencies in the literature (for the blue and gray dots). However, in this case Fig.\,\ref{fig:star3} shows a broad mass range rather than a sharp minimum, so the disagreement with the mass reported in the literature is acceptable.  \cite{Walczak2013} deduced a  linear relationship between the core overshooting, the mass, and the metalicity for the mode frequencies of HD\,866. Our PF-DNN results in Table\,\ref{tab:star3} are in agreement with their results when taking into account that linear relationship, if we keep in mind the factor between $\alpha_{\rm ov}$ and $f_{\rm ov}$. Indeed,  we find a lower mass and a lower metalicity than the values reported in \citet{Walczak2013}, but we report a higher core overshoot. All in all, we also find good agreement between our PF-DNN results and the literature values for this third star.

\subsection{The $\beta\,$Cep star HD\,29248}

The $\beta\,$Cep star HD\,29248 is one of the most intensively studied pulsators of its class. From the observational side, it was the first target of major months-long multisite multitechnique campaigns involving tens of astronomers worldwide \citep[see][for a summary]{Jerzy2005}.  That data led to the discovery of more modes than its radial mode and triplet, which were known prior to the
multisite campaign. Among the new modes are two high-order gravity modes. Recently, these results were confirmed by BRITE space photometry, which led by itself to yet a few more mode detections, bringing the total to ten low-order modes and seven high-order gravity modes \citep{Handler2017}. 

The star is also a relatively slow rotator, but the envelope rotation frequency does amount to $\sim\,0.12\,$d$^{-1}$, i.e., slightly faster than HD\,157056.
Again, we assume that we can ignore the effect of the Coriolis force on the oscillation frequencies and we work with the identified zonal modes for application to the PF-DNN. In summary, identification  of the zonal modes led to: 5.763264\,d$^{-1}$ for
$l=0, n_{pg}=+1$, 5.6373\,d$^{-1}$ for $l=1, n_{pg}=-1$, 6.2434\,d$^{-1}$ for $l=1, n_{pg}=+1$, and 7.9131\,d$^{-1}$ for $l=1, n_{pg}=+2$ \citep{DeRidder2004}.  HD\,29248 is the only $\beta\,$Cep pulsator so far with the detection and identification of three dipole pressure modes of consecutive radial order.

The seven additional modes without mode identification 
have frequencies 0.6870\,d$^{-1}$, 0.4307\,d$^{-1}$,
0.3894\,d$^{-1}$, 0.3395\,d$^{-1}$, 0.3242\,d$^{-1}$, 0.2951\,d$^{-1}$, and 0.2648\,d$^{-1}$ \citep{Handler2017}. While their radial orders $n_{pg}$ are not known, they must be decreasing in value (as these are gravity-mode frequencies whose $n_{pg}<0$). We assumed those to have mode degree $l=1$ or 2 and have added them to the PF-DNN.

\begin{figure}[h!]
\centering
\includegraphics[width=\textwidth]{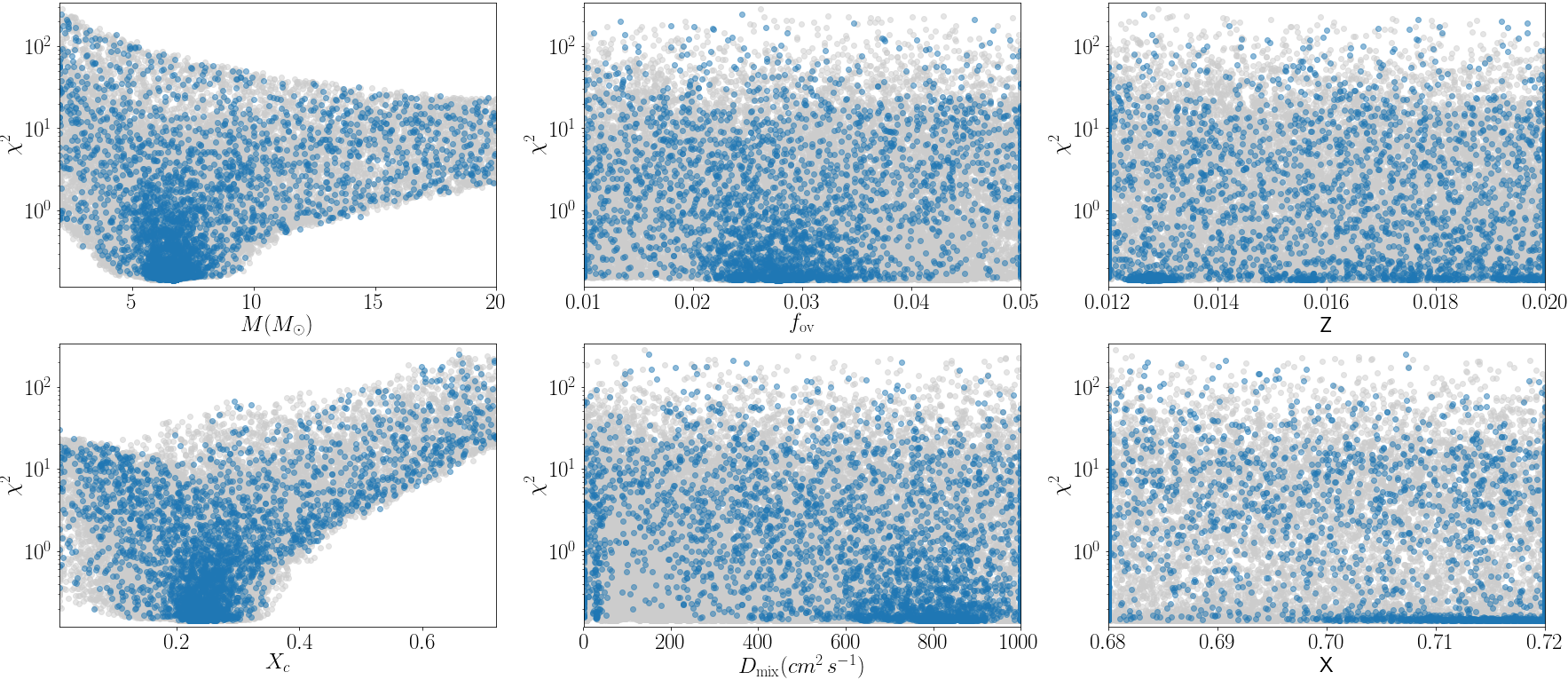}
\includegraphics[width=\textwidth]{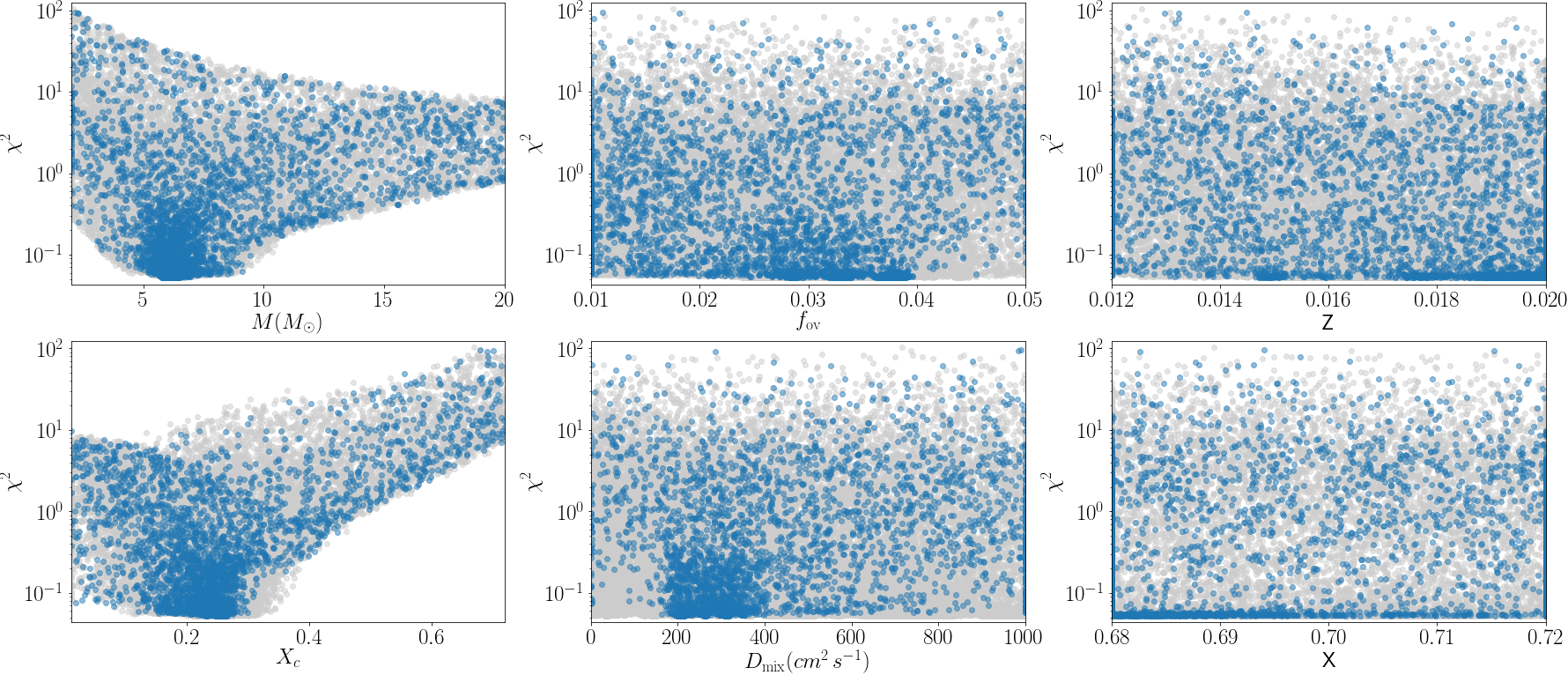}
\caption{Best PF-DNN fits for HD\,29248. Upper six panels: based on identified zonal low-order modes assuming the radial mode is the fundamental; lower six panels: case where the frequencies of seven unidentified high-order gravity modes were added to the identified modes.}
\label{fig:star4a}
\end{figure}

We applied the PF-DNN method twice: once for the case that the radial mode is the fundamental one with only identified modes and additionally by adding the unidentified gravity modes (see Fig.\,\ref{fig:star4a}). From both applications, it can be seen from Fig.\,\ref{fig:star4a} that the addition of the unidentified high-order modes does not help the PF-DNN. This is not surprising as, in general, feeding the PF-DNN with unidentified modes does not provide constraints on the stellar parameters.
The minima of the PF-DNN method for both cases (A: identified modes; B: non-identified mode frequencies added) are listed in Table\,\ref{tab:star4}, along with the results from seismic modeling in the literature based on the assumption of the fundamental radial mode \citep[][]{Ausseloos2004,Daszynska2017}. First of all, it can be deduced that the parameters of the two minima (here only given for the densest PF-DNN) are very similar. Furthermore, this is again a case where the value of the strict minimum for the mass is not very meaningful, because a similarly low $\chi^2$ is found for a mass range between some 6 to 9\,M$_\odot$ is revealed in Fig.\,\ref{fig:star4a}.
This broad mass range corresponds with the one found by two independent studies in the literature \citep{Ausseloos2004,Daszynska2017}. Irrespective of the mass range, the PF-DNN method finds HD\,29248 to be a rather evolved star, in agreement with the asteroseismic analyses by \citet{Ausseloos2004} and by \citet{Dziembowski2008}. We also deduce that a large amount of extra diffusive envelope mixing occurs, an aspect that was so far not studied for this star.

\begin{figure}[h!]
\centering
\includegraphics[width=\textwidth]{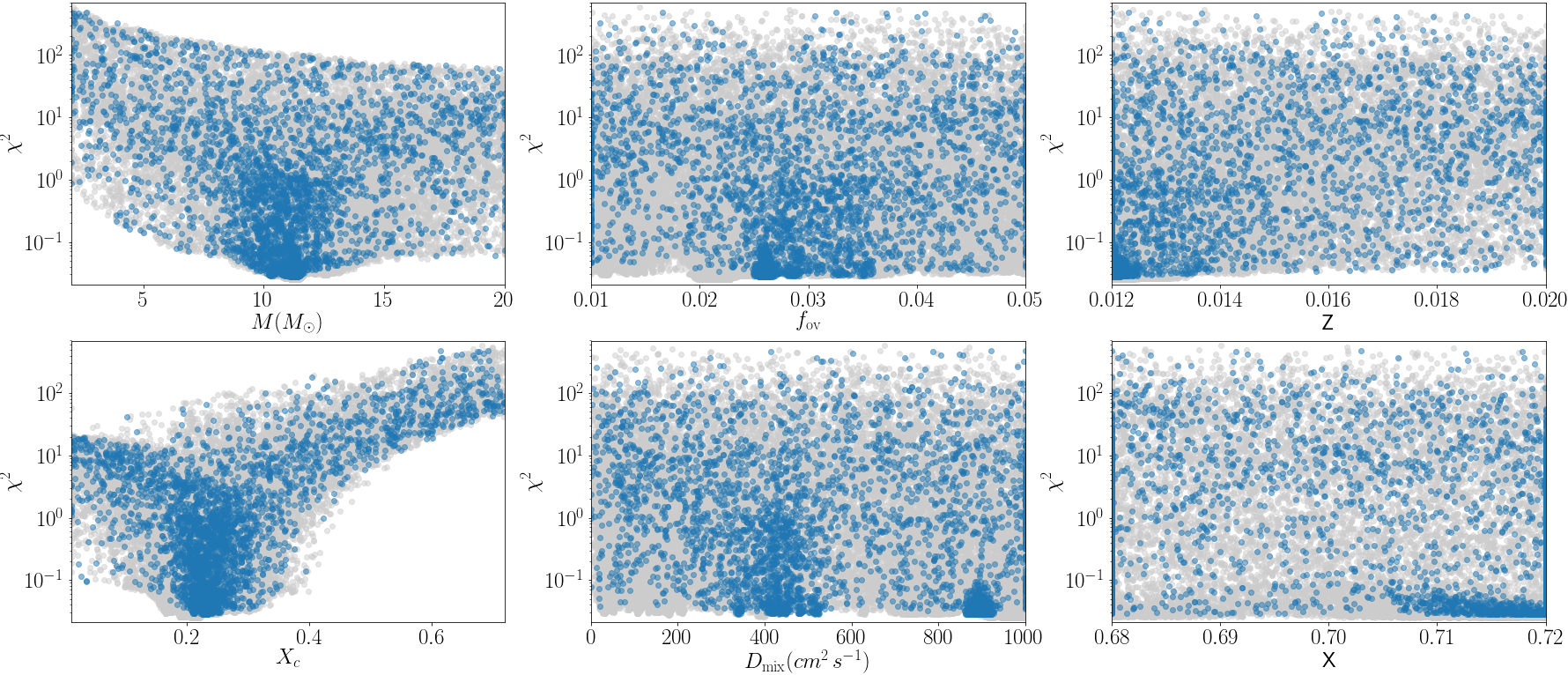}
\includegraphics[width=\textwidth]{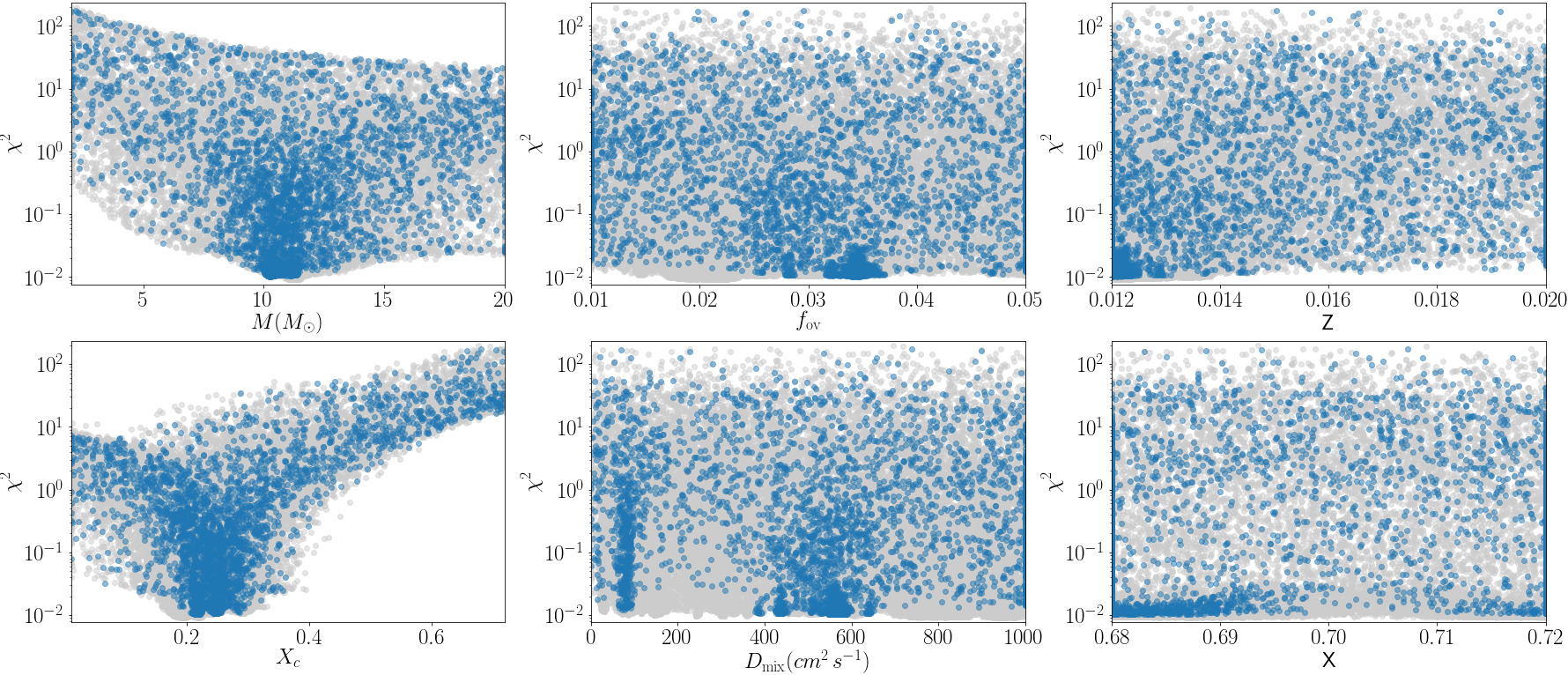}
\caption{Best PF-DNN fits for HD\,29248. Upper six panels: based on
  identified zonal low-order modes assuming the radial mode to be the first
  overtone; lower six panels: case where the frequencies of seven unidentified
  high-order gravity modes were added to the identified modes.}
\label{fig:star4b}
\end{figure}

\begin{table}[h!]
\centering
\caption{Two PF-DNN evaluations (indicated as Value A and Value B) for
  HD\,29248, once based on its four identified low-order modes assuming the
  radial fundamental mode and a second case where we added unidentified
  gravity-mode frequencies. Comparison with the literature results from two
  studies are given, both assuming that the radial mode is the fundamental.}
\label{tab:star4}
\begin{tabular}{lllll}
Parameter      & Value A (gray) & Value B (gray)   & \citet{Ausseloos2004} & \citet{Daszynska2017}   \\
\hline
$M$ (M$_\odot$)          & 6.56  & 6.55  & 7.83  & $\in [9.2,9.6]$  \\
$f_{\rm ov}$    & 0.0242         & 0.0255     & $\alpha_{\rm ov}=0.313$ & $\alpha_{\rm ov}\in [0.0,0.15]$ (fixed)  \\
$Z$               & 0.0195       & 0.0159 &0.0155   & $Z\in [0.015,0.0185]$     \\
$X_c$                    & 0.275  & 0.222& --  & --              \\
$D_{\rm mix} (cm^2\,s^{-1})$  & 770 & 1000       & --  & --         \\
$X$                                  & 0.716 &0.699        & 0.70 (fixed)   & 0.70 (fixed)     \\
\hline
\end{tabular}
\end{table}

For this star, we did an additional test by feeding the PF-DNN with the case where the dominant radial mode would be the first overtone, increasing as well the $n_{pg}$ of the modes with frequencies higher than the one of the radial mode. The outcome is shown in Fig.\,\ref{fig:star4b}. 
As can be seen in that plot, and as expected astrophysically, the solutions found by the PF-DNN now belong to higher mass models. Also for this case, the addition of extra unidentified gravity-mode frequencies does not change the morphology and best solutions found by the PF-DNN. This exercise illustrates that the PF-DNN is well capable to distinguish among model parameters when the overtone of the radial mode is different.

\subsection{The $\beta\,$Cep star HD\,214993}

Along with HD\,29248, the $\beta\,$Cep pulsator HD\,214993 has been the subject
of extensive months-long multicolor photometric and high-resolution
spectroscopic multisite campaigns. Its mode detection, mode identification and
forward asteroseismic modeling based on those ground-based data sets are
summarized in \citet{Handler2006,Dziembowski2008} and \citet{Desmet2009}.  In contrast to 
HD\,29248, there is no high-precision uninterrupted space photometry available
for HD\,214993, so we used the oscillation modes identified from the multisite
campaigns as input for the PF-DNN.

We made two applications for this star, based on the frequencies and mode
identification in Table\,8 of \citet{Desmet2009}, i.e., assuming the radial mode
to be the first overtone. These authors have deduced an average rotation
frequency of the star equal to 0.188$\pm$0.002\,d$^{-1}$, which is faster than
the other $\beta\,$Cep stars discussed so far.  This implies that the validity
of the SpaceInn grid of non-rotating models coupled to computations of the
oscillation modes while ignoring the Coriolis force might become sub-optimal for
HD\,214993.

\begin{figure}[h!]
\centering
\includegraphics[width=\textwidth]{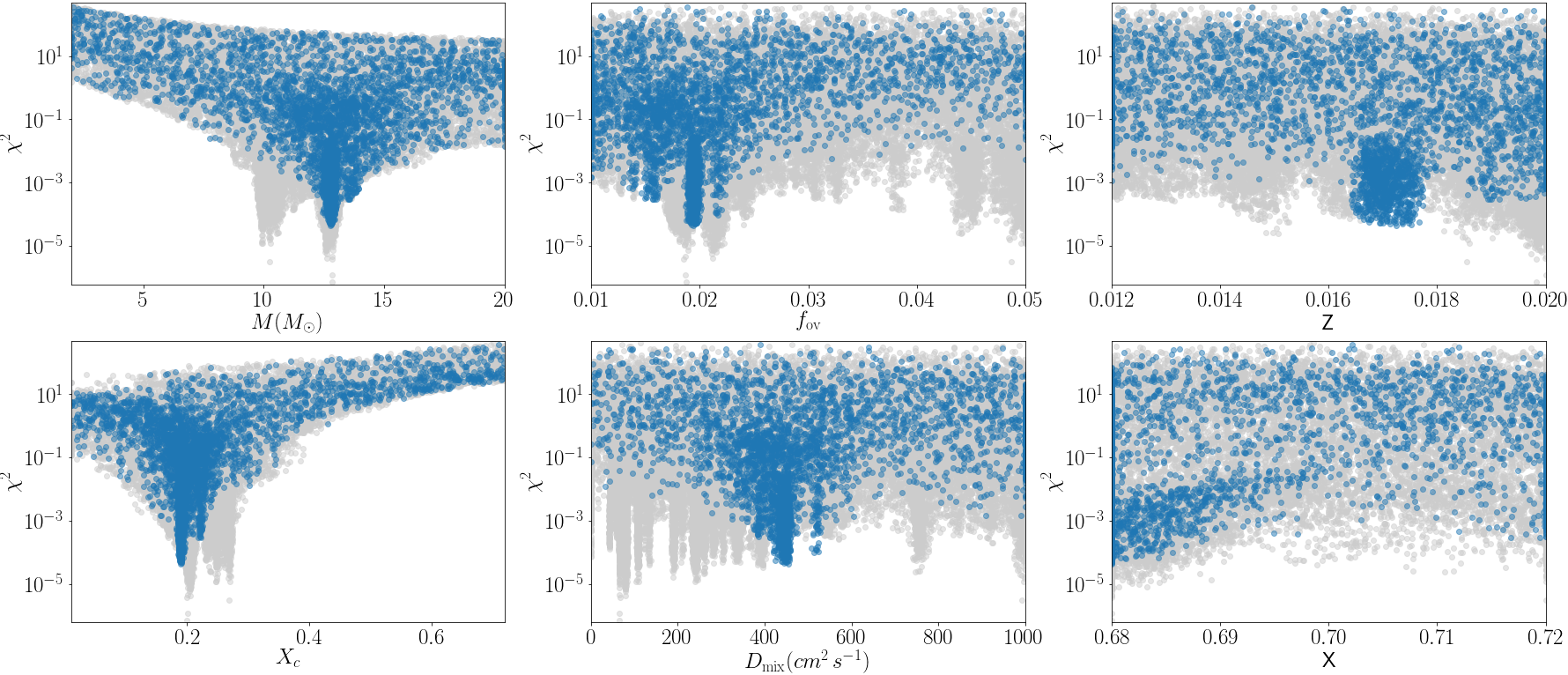}
\includegraphics[width=\textwidth]{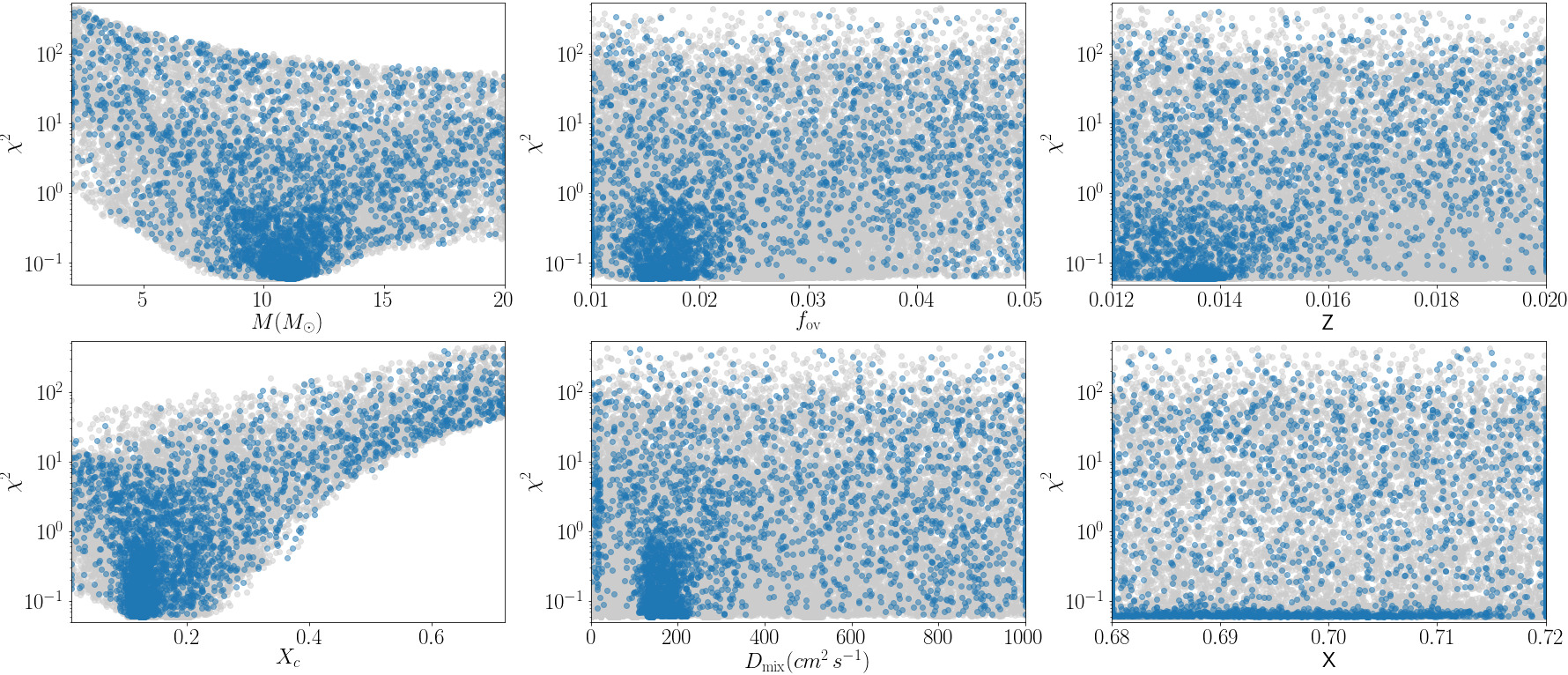}
\caption{PF-DNN method applied to HD\,214993, based on its three identified low-order 
zonal modes (upper six panels) and in the case where the additional low-order non-zonal modes are additionally fed into the PF-DNN.}
\label{fig:star5}
\end{figure}

We first performed a PF-DNN scan using only the three zonal modes, as
these are least affected by rotation.  It concerns the modes of frequencies
5.334224\,d$^{-1}$ with $l=0, n_{pg}=+2$, 5.066316\,d$^{-1}$ with
$l=1, n_{pg}=+1$, and 5.30912\,d$^{-1}$ with $l=2, n_{pg}=0$. The results of the
PF-DNN for this case are shown in the upper six panels of
Fig.\,\ref{fig:star5}. We recover good values for the mass, but find several
minima for the core overshooting.  Note that the ``manual'' forward modeling by
\citet{Desmet2009} could not pinpoint a value for $\alpha_{\rm ov}$.

If we instead add the
modes whose azimuthal order were identified as $m\neq 0$ or unidentified, namely 
5.490133\,d$^{-1}$ with   $l=2, n_{pg}= +1$,
4.241787\,d$^{-1}$ with   $l=2, n_{pg}=-2$, 
6.702318\,d$^{-1}$ with   $l=1, n_{pg}=+3$,
7.407162\,d$^{-1}$ with   $l=2, n_{pg}=+2$, and
5.84511\,d$^{-1}$ with     $l=1, n_{pg}=+2$ 
\citep[Table\,8,][]{Desmet2009}, then we find a lower mass, a lower metalicity
and a higher core overshoot (Fig.\,\ref{fig:star5}).  For this star, given it is
a moderate rather than a slow rotator, we see that the addition of frequencies
belonging to modes of $m\neq 0$, changes the morphology of solutions found by
the PF-DNN.  As a conclusion, the minima found by the PF-DNN
for both cases give meaningful results when compared with the modeling results
from \citet{Desmet2009}, even for the relatively high rotation rate of
0.188\,d$^{-1}$ for this pulsator.

\begin{table}[h!]
\centering
\caption{Best PF-DNN fit points for  compared with the asteroseismic solution
  in the literature. In the first PF-DNN case, we used the three identified zonal modes; in the 
  second case, we added five low-order non-zonal modes in the fit.}
\label{tab:star5}
\begin{tabular}{llll}
Parameter                    & PF-DNN Value A (gray) & PF-DNN Value B (gray)       & \citet{Desmet2009}    \\
\hline
$M$ (M$_\odot$)         & 12.8       &  10.2 &   $\in [10.2,14.4]$\\
$f_{\rm ov}$                 & 0.0187   & 0.0262       &    $\alpha_{\rm ov}\in [0.0,0.4]$   \\
$Z$                          & 0.0198   & 0.0132        &    0.015 (fixed)  \\
$X_c$                        & 0.200    & 0.174      &      $\in [0.13,0.21]$ \\
$D_{\rm mix} (cm^2\,s^{-1})$  & 66      & 279  &          --   \\
$X$                          & 0.68     & 0.72        &    0.72 (fixed)\\
\hline
\end{tabular}
\end{table}

\subsection{The Kepler SPB star KIC\,10526294}

We now turn our attention to the more difficult case of high-order gravity modes
in SPBs. These multiperiodic pulsators have been found since long from extensive
ground-based photometry \citep{Waelkens1991} and spectroscopy
\citet{Aerts1999}.  However, given that the periodicities of their high-order
gravity modes are half to several days, the beating patterns of the overall
effect of the multiperiodic oscillations cover many years. It is therefore needed
to have time-series data with a very long time base. Despite extensive
ground-based campaigns covering many years, these data turned out to be of too low duty cycle and
led to too few modes to perform forward asteroseismic modeling
\citep[cf.,][]{DeCatAerts2002}.  This capacity had to await the 4-year {\it Kepler\/}
photometry, which revealed period-spacing patterns of the gravity modes. Such
patterns allow to identify the mode degrees $l$ and estimate the radial orders
$n_{pg}$. 

\citet{Moravveji2015} performed the first forward seismic modeling of an SPB. Their target was 
the B8.3V star KIC\,10526294, whose frequencies were derived from a 4-year
long {\it Kepler\/} light curve by \citet{Papics2014}.  The modeling was based
on 19 detected dipole zonal modes with consecutive radial order (listed in Table\,1 of \citet{Moravveji2015},
and omitted here for brevity). These 19 modes reveal an average period spacing of 5,400\,s.
Their forward modeling led to identification of
the radial orders $n_{pg}$ ranging from -14 to -32.  Based on the best
equilibrium models, \citet[][]{Triana2015} subsequently determined the star's
rotation profile throughout its interior from inversion methods applied to the
rotationally split triplet frequencies. They found a counter-rotating profile
that deviates only slightly from rigidity inside the star. KIC\,10526294 was
found to rotate exceptionally slowly for a B star, with an average rotation rate
of only 0.0054\,d$^{-1}$. 
This star is therefore the ideal and so far only case to test the capacity
of our PF-DNN method for stars that only exhibit high-order gravity modes and no pressure modes. Indeed, all other {\it Kepler\/} or BRITE SPBs with period spacing patterns have fast rotation 
\citep{Papics2017,Kallinger2017,Szewczuk2018}
and cannot be treated with the SpaceInn grid.

\begin{figure}[h!]
\centering
  \includegraphics[width=\textwidth]{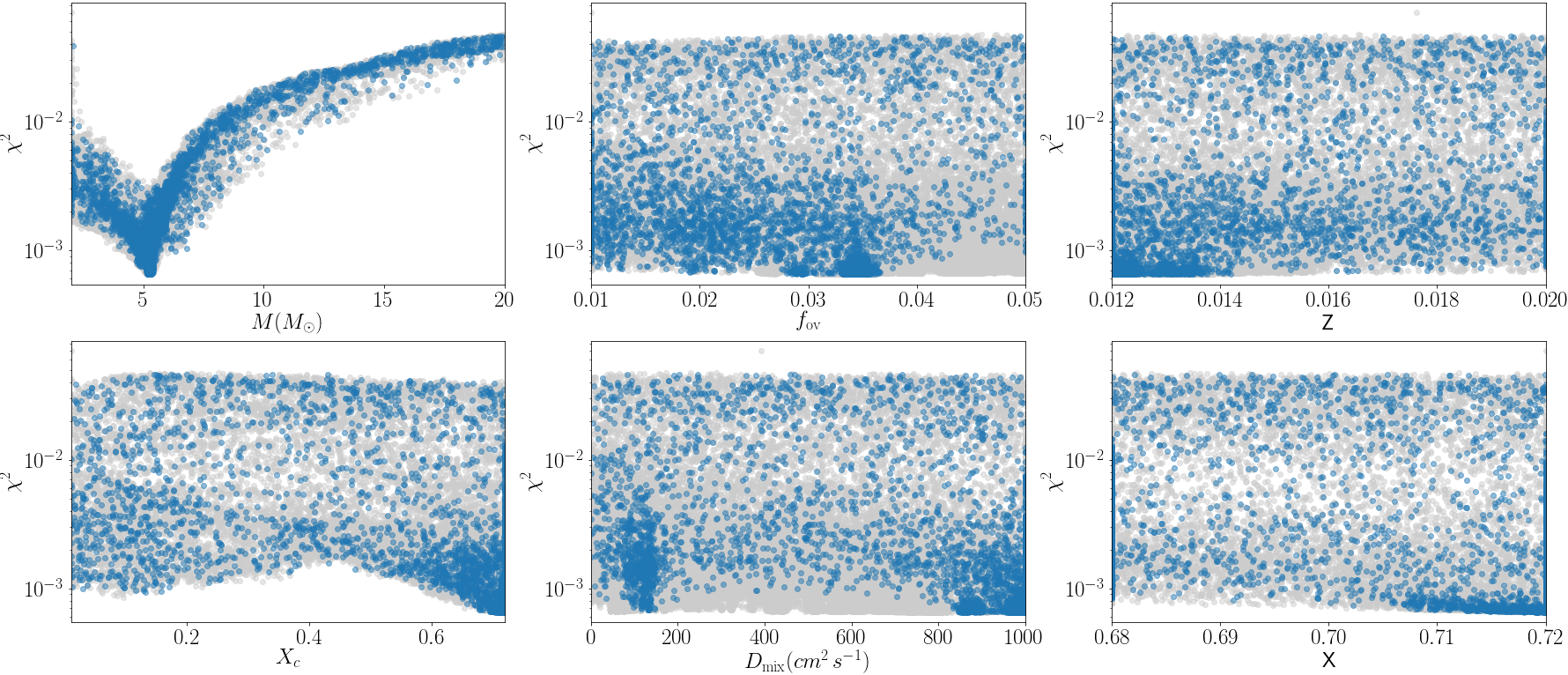}
\caption{PF-DNN method applied to KIC\,, based on its 19 zonal dipole gravity 
modes with consecutive radial order.}
\label{fig:star6}
\end{figure}

\begin{table}[h!]
\centering
\caption{Best PF-DNN fit point for KIC\,10526294 compared with the latest asteroseismic solution
  in the literature.}
\label{tab:star6}
\begin{tabular}{llll}
Parameter                    & PF-DNN Value (gray) & PF-DNN Value (blue)      & Paper\,I    \\
\hline
$M$ (M$_\odot$                    & 5.25   & 5.26    & 3.24     \\
$f_{\rm ov}$                 & 0.0348 & 0.0345   & 0.017             \\
$Z$                          & 0.0120 & 0.0120  & 0.017   \\
$X_c$                        & 0.712  & 0.712  & 0.644             \\
$D_{\rm mix} (cm^2\,s^{-1})$  & 798   & 996 & 178      \\
$X$                          & 0.72   & 0.72 & 0.71 (fixed)\\
\hline
\end{tabular}
\end{table}

We fed the PF-DNN with the 19 zonal dipole mode frequencies and opted not to pinpoint the $n_{pg}$ values to those found in \cite{Moravveji2015}, but rather set the constraint that all the 19 modes must be of consecutive radial order. This is the best approach given that the mode spectrum for high-order gravity modes is very dense and labeling of $n_{pg}$ is sometimes ambiguous for close frequency values. The outcome can be seen in Fig.\,\ref{fig:star6} and the minima found by the two versions of the PF-DNN are listed in Table\,\ref{tab:star6}. We find a sharp minimum for the mass, but at a value about 2\,M$_\odot$ higher than the one obtained by a re-analysis of the forward modeling as presented in Paper\,I. However, the PF-DNN also finds a twice as high core overshoot parameter. Keeping in mind the $(M, f_{\rm ov})$ degeneracy \citep[cf.\ Fig.\,5 in][]{Moravveji2015}, it is expected that a higher mass must be accompanied by a higher core overshoot. Moreover, the lower metalicity $Z$ found by the PF-DNN must also be accompanied by a higher mass. Finally, the PF-DNN's minima occur at the ZAMS (highest initial hydrogen ($X$) value), i.e., a border of the SpaceInn grid. Taking the degeneracies among the 
$(M, f_{\rm ov}, Z, D_{\rm mix})$ as in Fig.\,5 of \citet{Moravveji2015}, the PF-DNN solution is well understood but illustrates that its application to the high-order gravity modes of SPBs is less powerful than to the low-order pressure modes of $\beta\,$Cep stars.

\subsection{The CoRoT SPB HD\,50230}

The B3V star HD\,50230 was the first main-sequence star for which a period
spacing pattern of high-order gravity modes was discovered \citep{Degroote2010}.
This discovery was based on 5 months of CoRoT space photometry and led to an
average spacing of $\sim 9,400\,$s from eight modes with consecutive radial
order. From this average spacing, \citet{Degroote2010} derived a star of
$\sim 7\,$M$_\odot$, which is relatively far evolved in its core-hydrogen
burning stage ($X_c < 0.3$). The small and periodic deviation from the average
period spacing made the authors conclude that convective penetration cannot be
the only mixing phenomenon in the radiative envelope, as this would lead to a
sharp $\mu$-gradient in the near-core region and hence strong mode
trapping. This is not observed for this star, hence the conclusion that
diffusive envelope mixing must occur. A rough estimate of $D_{\rm mix}$ between
1,000 and 10,000\,cm$^2$\,s$^{-1}$ was suggested.

Regrettably, the CoRoT data is only five months in duration and this turns out
to be a limitation for forward asteroseismic modeling (Paper\,I), unless one can
get independent additional information on the  degree $l$ of at least one mode
\citep{Buysschaert2018}.  This extra information is not available for HD\,50230. Moreover, \citet{Degroote2012} discovered HD\,50230 to
be part of a spectroscopic binary with a wide orbit, which they did not manage
to cover.

\begin{figure}[h!]
\centering
  \includegraphics[width=\textwidth]{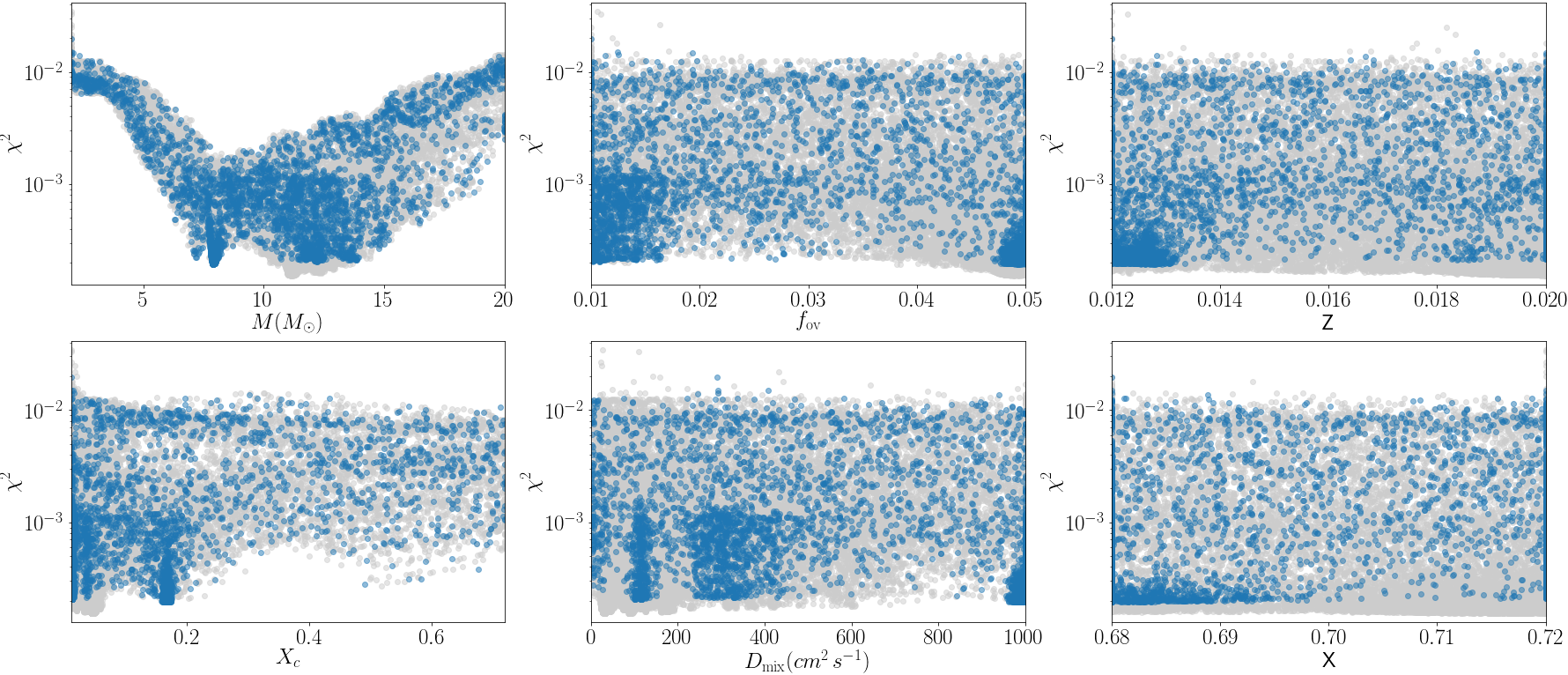}
\caption{PF-DNN method applied to HD\,50230, based on its high-order gravity  modes.}
\label{fig:star7}
\end{figure}

\begin{table}[h!]
\centering
\caption{Best PF-DNN fit point for HD\,50230. For this star, there is no
  asteroseismic 
modeling done yet, but we list the rough parameter evaluation.}
\label{tab:star7}
\begin{tabular}{llll}
Parameter                    & PF-DNN Value (gray) & PF-DNN Value (blue)    & \citet{Degroote2010}      \\
\hline
$M$ (M$_\odot$)                    & 11.12  & 7.88   & $\sim\,7$          \\
$f_{\rm ov}$                 & 0.0494 & 0.0494  & $\alpha_{\rm ov}\in [0.2,0.5]$              \\
$Z$                          & 0.02  & 0.0129 & 0.02 (fixed)               \\
$X_c$                        & 0.0453  & 0.173   & $< 0.3$            \\
$D_{\rm mix} (cm^2\,s^{-1})$  & 32    & 1000 & --           \\
$X$                          & 0.716  & 0.68  & 0.71 (fixed) \\
\hline
\end{tabular}
\end{table}

Here, we analysed HD\,50230 with our PF-DNN and find an excellent solution from the less dense network (cf.\ Fig.\,\ref{fig:star7}). 
This is due to the fact that the denser one has a broad deeper minimum at a too high mass, corresponding with very old stellar models close to the end of the core-hydrogen burning. The deepest minimum of the less dense PF-DNN corresponds to very plausible parameters for this star and brings the first estimate of its large core overshooting and envelope mixing. Here, it is the average period spacing value that allows us to eliminate the too high mass solution. Without that information, any user would have to perform detailed forward modeling, keeping the two different minima for the mass in any more detailed follow-up study.


\section{Conclusion and Future Prospects}
\label{sec:conclusion}

We proposed a deep learning method as a starting point for the forward
asteroseismic modeling of stars with coherent oscillation modes. A deep neural
network was trained on an asteroseismological model grid containing more than 60
million theoretically predicted oscillation modes. Training the neural
  network has to be done once and then is applicable to all stars within the
  grid of the training set. This took around six hours on an Nvidia GTX 1080.
By using a genetic algorithm over this trained neural network to find
  the maximum likelihood one can successfully find appropriate regions for the
stellar parameters in a 6D space in less than a minute, relying only on detected
and (partially) identified oscillation-mode frequencies. Our PF-DNN
  method was tested and applied to seven stars that were already modeled
``manually'' in the past. PF-DNN retrieved appropriate intervals for
the parameters of those pulsators compared to the results available in the
literature, keeping in mind known degeneracies between the stellar parameters.

Even though the DNN prediction is not perfect, it does reveal the regions in
multi-D model space upon which to focus for more detailed screening. This method
is therefore suitable as a first sampling strategy to constrain the possible
complex parameter space. Indeed, the results of the network can be used as a
seed point for more advanced applications of forward asteroseismic modeling
following Paper\,I, for which the detailed and CPU-intensive computation of
dedicated refined stellar model and pulsation grids as in, e.g.,
\citet{Moravveji2015} is necessary. Application of our method greatly helps the
strategy and planning of such detailed grid computations after initial rough
screening of the multi-D parameter space.

In a next step, we will generalize our PF-DNN method to rotating
pulsating stars for which the Coriolis force cannot be neglected as we did
here. This requires the addition of at least one dimension in the
PF-DNN, namely the rotation frequency of the star. Accordingly, the
shifts of the oscillation frequencies due to the Coriolis force must be computed
for each of the modes in each of the models of the grid.  While this is a
serious complication, it also allows many more applications, given that most
pulsators rotate at a frequency that cannot be neglected when fitting the
measured non-zonal oscillation modes. Once our PF-DNN method is
generalized to treat this case, it will be an appropriate starting tool to apply
forward asteroseismic modeling with the statistical methodology of parameter
estimation and model selection described in Paper\,I. This method by itself
relies on a two-step approach: estimation of the rotation frequency of the star
from detected multiplets of coherent modes
\citep[e.g.,][]{Kurtz2014,Papics2014,Saio2015} and/or from period spacing
patterns of high-order gravity modes
\citep{VanReeth2016,Papics2017,Ouazzani2017}, followed by frequency fitting of
all modes with identified degree $l$ and azimuthal order $m$ to within the
measurement errors of the frequencies. Such a combined initial PF-DNN
  method followed by a statistical estimation and model selection approach
offers an optimal way of forward asteroseismic modeling to be applied to many
hundreds of pressure- and gravity-mode pulsators for which appropriate {\it
  Kepler}, TESS \citep{Ricker2016}, and PLATO \citep{Rauer2014} data are or will
become available in the near future.

In the next step, the PF-DNN method itself will be improved as well. A
  particular downside of the current setup is that the method does not contain
  uncertainty information. One can improve this by changing the DNN to a
  Bayesian DNN including aleatoric and epistemic uncertainties following the
  procedure of \citet{DBLP:journals/corr/KendallG17}. In that way, the maximum
  likelihood evaluation can be replaced with nested sampling \citep[described
  in][]{doi:10.1063/1.1835238} to search the parameter space in a Bayesian
  context. This will provide not only a prediction of the model parameter ranges
  but contain uncertainty information as well. Such a procedure can then be
  followed by manual analyses zooming in on the selected region of interest, by
  allowing for various structural descriptions of core overshooting
  $D_{\rm ov}(r)$ and mixing $D_{\rm mix}(r)$ to derive their shape in the
  stellar interior, using the statistical methodology for stellar model
  performance comparison outlined as Problem Set \,4 in \citet{Paper-I}. 

\acknowledgments We express our sincere appreciation to the developer teams of
the MESA and GYRE codes for making their software publicly available and for
their continuous efforts to guide and educate the astronomical community in
their use.  The computation of the stellar model grid used in this paper was
done by Dr.\,Valentina Schmid and was funded by the European Community's Seventh
Framework Programme FP7-SPACE-2011-1 (project number 312844: SpaceInn) and by
the VSC (Flemish Supercomputer Center) funded by the Flemish Government –
department EWI.  We are grateful to Valentina for her dedication to transport
and deliver the hard-disc with the model grid (carefully packed among a lot of
other stuff) from Leuven to Nijmegen. Both referees are acknowledged for their
useful comments and requests to add more analysis details to our manuscript,
resulting in an improved presentation. The research leading to these results
has received funding from the European Research Council (ERC) under the European
Union's Horizon 2020 research and innovation programme (grant agreement
N$^\circ$670519: MAMSIE).

\software{MESA \citep{Paxton2011,Paxton2013,Paxton2015,Paxton2018}, 
GYRE \citep{Townsend2013,Townsend2018}}

\newpage

\bibliographystyle{aasjournal}
\bibliography{bibliography}

\end{document}